\documentclass[twocolumn]{emulateapj}  
\usepackage{bm}
\usepackage{csvsimple}
\usepackage{arydshln}
\usepackage{epstopdf}
\usepackage{ulem}
\usepackage{amssymb}
\usepackage{times}
\usepackage{graphicx}
\usepackage[usenames,dvipsnames]{pstricks}
\usepackage{psfrag}
\usepackage{natbib}
\usepackage[colorlinks=true,linkcolor=blue,citecolor=blue]{hyperref}
\usepackage[cp1251]{inputenc}
\def\araa{ARA\&A}%% Annual Review of Astron and Astrophysf
\def\apj{ApJ}%% Astrophysical Journal
\def\apjl{ApJ}%% Astrophysical Journal, Letters
\def\aap{A\&A}%% Astronomy and Astrophysics
\def\mnras{MNRAS}%% Monthly Notices of the RAS
\def\nat{Nature}%% Nature
\def\physrep{Phys.~Rep.}%% Physics Reports
\newcommand{\be}{\begin{equation}}
\newcommand{\ee}{\end{equation}}
\newcommand{\bary}{\begin{eqnarray}}
\newcommand{\eary}{\end{eqnarray}}

\begin{document}
%\linenumbers

%------------------------------------------------------------------------------------------------------------
% Title
%------------------------------------------------------------------------------------------------------------
%Interpretation of tand  a break synchrotron
\title{GRB Fermi-LAT afterglows: explaining flares, breaks, and energetic photons}
%------------------------------------------------------------------------------------------------------------
% Authors
%------------------------------------------------------------------------------------------------------------
\author{N.~ Fraija\altaffilmark{1$\dagger$}, T. Laskar\altaffilmark{2}, S. Dichiara\altaffilmark{3,4},   P. Beniamini\altaffilmark{5},  R. Barniol Duran\altaffilmark{6}, M.G. Dainotti\altaffilmark{7,8, 9,10} and R. L. Becerra\altaffilmark{1}}
\affil{$^1$Instituto de Astronom\'ia, Universidad Nacional Aut\'{o}noma de M\'{e}xico, Apdo. Postal 70-264, Cd. Universitaria, Ciudad de M\'{e}xico 04510}
\affil{$^2$ Department of Physics, University of Bath, Claverton Down, Bath, BA2 7AY, United Kingdom}
\affil{$^3$ Department of Astronomy, University of Maryland, College Park, MD 20742-4111, USA}
\affil{$^4$ Astrophysics Science Division, NASA Goddard Space Flight Center, 8800 Greenbelt Rd, Greenbelt, MD 20771, USA}
\affil{$^5$ TAPIR, Mailcode 350-17, California Institute of Technology, Pasadena, CA 91125, USA}
\affil{$^6$ Department of Physics and Astronomy, California State University, Sacramento, 6000 J Street, Sacramento, CA 95819-6041, USA}
\affil{$^7$  Physics Department, Stanford University, 382 Via Pueblo Mall, Stanford, USA}
\affil{$^8$ Space Science Institute, Boulder, Colorado}
\affil{$^9$ Obserwatorium Astronomiczne, Uniwersytet Jagiello\'nski, ul. Orla 171, 31-501  Krak\'ow, Poland}
\affil{$^{10}$ Interdisciplinary Theoretical \& Mathematical Science Program, RIKEN (iTHEMS), 2-1 Hirosawa, Wako, Saitama, Japan 351-0198}
\email[$\dagger$ ]{nifraija@astro.unam.mx}
\begin{abstract}
The Fermi-LAT collaboration presented the second gamma-ray burst (GRB) catalog covering its first 10 years of operations. A significant fraction of afterglow-phase light curves in this catalog cannot be explained by the closure relations of the standard synchrotron forward-shock  model, suggesting that there could be an important contribution from another process.   In view of the above,  we derive the synchrotron self-Compton (SSC) light curves from the reverse shock in the thick- and thin-shell regime for a uniform-density medium.  We show that this emission could explain the GeV flares exhibited in some LAT light curves.   Additionally, we demonstrate that the passage of the forward shock synchrotron cooling break through the LAT band from jets expanding in a uniform-density environment may be responsible for the late time ($\approx10^2$~s) steepening of LAT GRB afterglow light curves.   As a particular case,  we model the LAT light curve of GRB 160509A that exhibited a GeV flare together with a break in the long-lasting emission, and also two very high energy photons with energies of 51.9 and 41.5 GeV  observed 76.5 and 242 s after the onset of the burst, respectively.   Constraining the microphysical parameters and the circumburst density from the afterglow observations,  we show that the GeV flare is consistent with a SSC reverse-shock model,  the break in the long-lasting emission with the passage of the synchrotron cooling break through the Fermi-LAT band and the very energetic photons with SSC emission from the forward shock when the outflow carries a significant magnetic field ($R_{\rm B} \simeq 30$) and it decelerates in a uniform-density medium with a very low density ($n=4.554^{+1.128}_{-1.121}\times 10^{-4}\,{\rm cm^{-3}}$). 
\end{abstract}
%
%------------------------------------------------------------------------------------------------------------
% Keywords
%------------------------------------------------------------------------------------------------------------

%------------------------------------------------------------------------------------------------------------
% Section 1: Introduction 
%
%------------------------------------------------------------------------------------------------------------
\section{Introduction}
Gamma-ray bursts (GRBs) are  the most energetic astrophysical sources in the universe. These events exhibit intense and nonuniform gamma-ray flashes created by inhomogeneities  within the ultra-relativistic outflows \citep{2004IJMPA..19.2385Z}.    The temporal and spectral features inferred from the early and late emissions, usually known as prompt-to-afterglow, respectively, can be interpreted within the context of the fireball model \citep[e.g., see][]{1978MNRAS.183..359C}.  This involves a relativistic outflow that moves into the circumstellar medium and generates an outgoing (forward) shock and a reverse shock (RS) that propagates back into the outflow.  Electrons accelerated in the forward shock (FS), and RS are cooled down principally by synchrotron and synchrotron-self Compton (SSC) emission.  The synchrotron emission from the FS region is predicted to produce long-lasting afterglow emission \citep{1998ApJ...497L..17S} that can extend to $>100~$MeV energies \citep{2009MNRAS.400L..75K,2010MNRAS.409..226K}. Synchrotron emission from the RS is expected to generate an optical flash or an X-ray flare \citep{2007ApJ...655..973K, 2007ApJ...655..391K, 2000ApJ...545..807K, 2018ApJ...859...70F, 2019ApJ...881...12B, 2020arXiv200404179A}.  If no new electrons are injected, then the material at the RS cools adiabatically.  Synchrotron photons at the RS region can be up-scattered by the same electron population in order to describe the gamma-ray flares \citep{2007ApJ...655..973K, 2015ApJ...804..105F, 2016ApJ...818..190F, 2016ApJ...831...22F}.    The prompt-to-afterglow transition  is usually observed in the X-ray light curves as a steep decay interpreted  as high-latitude emission \citep{2000ApJ...541L..51K, 2006ApJ...642..389N, 2010MNRAS.403.1296W} or in some cases as the X-ray flares / optical flashes described in terms of the synchrotron  radiation and SSC emission from an early afterglow \citep[e.g., see][]{2003ApJ...595..950Z, 2003ApJ...597..455K, 2005ApJ...628..315Z, 2007ApJ...655..973K}. 

Recently, \cite{Ajello_2019} presented the second Fermi-LAT (Large Area Telescope) GRB catalog, covering the first 10 years of operations (from 2008 to 2018 August 4). The catalog comprises 169 GRBs with high-energy emission at $\ge100$~ MeV, including 29 GRBs with prompt emission extending to $\ge10$~GeV.  A large number (86) of LAT GRBs exhibited temporally-extended (hereafter called ``long-lasting'') emission.  A subset of these events (21 GRBs) exhibited a break in the LAT light curve between 63 and 1250~s.  Although the long-lasting emission is usually interpreted as synchrotron radiation from external FSs \citep{2009MNRAS.400L..75K,2010MNRAS.409..226K}, not all the LAT light curves satisfy the relation between power-law (PL) temporal and spectral indices or the closure relations that are expected in case the FS dominates the emission \citep{1998ApJ...497L..17S}.  Furthermore, there is evidence that the entirety of the GeV light curve for several bursts cannot be adequately explained as originating in FS synchrotron radiation alone \citep{2011MNRAS.415...77M}.  In addition, half the bursts with long-lasting LAT emission exhibit light curves that peak before the prompt emission ends. 

In bursts with high isotropic-equivalent prompt $\gamma$-ray energy ($E_{\rm iso} \gtrsim 10^{53}\,{\rm erg}$) such as GRB 080916C, GRB 090510A, GRB 090902B, GRB 090926A, GRB 110731A and GRB 130427A \citep[e.g., see][]{2009Sci...323.1688A, 2009ApJ...706L.138A, 2010ApJ...716.1178A, 2011ApJ...729..114A, 2013ApJ...763...71A, 2014Sci...343...42A} the LAT light curves  exhibited a short and bright peak (hereafter called GeV flare).   For instance,   GRB 160509A  exhibited a GeV flare peaking at $\sim$ 20 s, followed by a long-lasting emission with a break at $\sim$ 300 s. This burst exhibited  the second highest-energy photon reported by the second Fermi-LAT GRB catalog,  a 52 GeV event  arrived 77 s after the onset of the burst.  

To interpret GeV flares in the LAT  light curves,  we derive the SSC light curves from the RS.  We also investigate the break exhibited at hundreds of seconds using the synchrotron FS model.   As a particular case of the thick-shell regime, we describe the  LAT observations (the GeV flare and the break in the long-lasting emission) of GRB 160509A and constrain the model parameters from the afterglow observations.  We interpret the very energetic photons detected in the SSC framework from the FS.  As an example of the thin-shell regime, we calculate the SSC emission from the optical flash displayed in GRB 180418A.

The paper is organized as follows. In Section 2, we derive the SSC light curves from the RS in the thick- and thin-shell regime and analyze the break exhibited in the long-lasting emission in the synchrotron FS framework.  In Section 3, we model the LAT observations of GRB 160509A and discuss the implications of the results.    In Section 4, we estimate the SSC light curves in the thin-shell regime for GRB 180418A, and in Section 5, we summarize.  The convention $Q_{\rm x}=Q/10^{\rm x}$  in c.g.s. units will be adopted throughout this paper.  

\section{Connection between theory and  Fermi LAT observations}\label{sec2}

\subsection{SSC light curves from RS}\label{ssc_lightcurves}

A RS propagating into the GRB jet  (a ``shell") is produced  when the relativistic ejecta interacts with the external medium \citep{1997ApJ...476..232M, 1999ApJ...517L.109S, 2000ApJ...542..819K}.   The dynamics of the RS can be described through the Sedov length \citep{1995ApJ...455L.143S}

\be\label{ell}
\ell=\left(\frac{3E_{\rm K}}{4\pi m_pc^2\, n }\right)^{\frac{1}{3}}\,,
\ee

and the bulk Lorentz factor during the deceleration phase 

\bary\label{gamma}
\Gamma&=&\left(\frac{\ell}{\Delta} \right)^\frac38,\cr
&=&\left(\frac{3 E_{\rm K} (1+z)^3 }{32\pi m_p c^5 n}\right)^\frac18\,t_{\rm x}^{-\frac38}\,, 
\eary

where $\Delta=2c (1+z)^{-1}\,t_{\rm x}$ is the observed width of the shell given by the RS crossing time $t_{\rm x}$,  $n$ is the density of the circumburst medium (assumed uniform),  $E_{\rm K}$ is the isotropic equivalent kinetic energy, $c$ is the speed of light and  $m_p$ is the proton mass.  Taking into account the duration of the burst ($T_{90}$\footnote{The choice of $T_{90}$ can be indicative of the duration of the burst, but this is not the only viable choice \citep{2007ApJ...662.1093W, 2011MNRAS.418.2202D, Ajello_2019}}),  it is possible to define two regimes:  the thick-shell regime for $\Gamma_c < \Gamma$ (corresponding to $t_{\rm x} \lesssim T_{90}$) and the thin-shell regime for $\Gamma < \Gamma_c$ ($T_{90} <   t_{\rm x}$), where the critical Lorentz factor is defined by 

\be
\Gamma_c\equiv  \left(\frac{3E_{\rm K}\,(1+z)^3}{32\pi m_p c^5\, n}\right)^{\frac{1}{8}}  T_{90}^{-\frac38}\,.
\ee

The RS accelerates electrons in the shell to relativistic energies, which radiate photons via synchrotron and SSC processes. The synchrotron process has been widely explored and discussed in \cite{2000ApJ...545..807K}. We expand the established RS synchrotron formalism to include the expected contribution of SSC emission in a uniform-density environment in both thick- and thin-shell limits.\\

\subsubsection{The thick-shell regime ($\Gamma_{\rm c}< \Gamma$)}
In the thick-shell regime, the RS becomes ultra-relativistic and therefore, can decelerate the shell substantially.   In this case, the shock crossing time is less than the duration of the burst ($ t_{\rm x} \lesssim T_{90}$).  To derive the SSC light curves for the thick-shell regime, we describe the dynamics before and after the shock crossing time, separately. 

\paragraph{Before the shock crossing time ($t < t_{\rm x}$)}  At $t<t_{\rm x}$, the Lorentz factor of the fluid behind the RS evolves as $\gamma_3\propto t^{-1/4}$ due to adiabatic expansion,\footnote{During RS crossing, the RS and FS divide the system into four regions (1) unshocked ISM (2) shocked ISM (3) shocked ejecta, and (4) unshocked ejecta, respectively.} where $t$ is the lab-frame time. During this period and in the same region, the post-shock magnetic field,\footnote{Primes correspond to the quantities in a comoving frame.} $B^{\prime}\propto t^{-1/4}$, and the Lorentz factor of the lowest-energy electrons, $\gamma_{\rm m,r} \propto t^{1/4}$. Electrons above $\gamma_{\rm c,r}\propto t^{-1/4}$ cool efficiently by synchrotron emission. The corresponding synchrotron break frequencies at the RS region evolve as $\nu_{\rm m,r}^{\rm syn}\propto t^0$ and $\nu^{\rm syn}_{\rm c, r}\propto t^{-1}$, while the spectral peak flux density, $F^{\rm syn}_{\rm max,r }\propto t^\frac12$ \citep{2000ApJ...545..807K}.  Hereafter, the sub-index ``r" refers to the derived quantities in the RS. \\

Electrons accelerated  by the RS can up-scatter synchrotron photons, yielding an SSC spectrum characterized by the break frequencies $h\nu^{\rm ssc}_{\rm i,r}\sim\gamma^2_{\rm i,r} h\nu^{\rm syn}_{\rm i,r}$ with ${\rm i=m}$ and ${\rm c}$.   The flux of SSC emission can be computed as $F^{\rm ssc}_{\rm max}\sim\, \frac43\frac{p-2}{p-1} \sigma_T n R \,F^{\rm syn}_{\rm max}$, where $R$ is the deceleration radius and $\sigma_T$ the Thompson cross section \citep{2000ApJ...545..807K}.\\

Based on the foregoing, the spectral breaks and the maximum flux of SSC emission can be written as\footnote{$\Gamma_3=\Gamma/10^3$ is the bulk Lorentz factor at the deceleration time, given by equation (2), and is not to be confused with $\gamma_3$, the fluid Lorentz factor behind the shock.} 
{\small
\bary\label{ssc_before}
h\nu^{\rm ssc}_{\rm m, r}&\simeq& 6.1\times 10^2\,{\rm eV}\,\left(1+z\right)^{-1}\,\epsilon^4_{\rm e_r,-1} \epsilon^{\frac12}_{\rm B_r,-1} n^\frac34_{-4} \Gamma^4_{3}\, \Delta^\frac14_{12} \, \cr
&& \hspace{5.5cm} \times\,E^{-\frac14}_{\rm K,53} t^{\frac12}_1\,,\cr
h\nu^{\rm ssc}_{\rm c, r}&\simeq&  2.1\times 10^3\,{\rm TeV} \left(1+z\right)^{3}\, \left(1+Y_{\rm r} \right)^{-4}\,\epsilon^{-\frac72}_{\rm B_r,-1} n^{-\frac94}_{-4} \Delta^\frac54_{12}\,E^{-\frac54}_{\rm K,53},\cr
&& \hspace{6.4cm} \times \, t^{-\frac32}_1\,\cr
F^{\rm ssc}_{\rm max,r} &\simeq&  3.6\times 10^{-6}\,{\rm m Jy}\,\,  \epsilon^{\frac12}_{\rm B_r,-1}\,n_{-4} \, \Gamma^{-1}_{3}\, \Delta^{-\frac32}_{12}\, d^{-2}_{\rm z,28}\,E^{\frac32}_{\rm K,53} t_1\,,
\eary
}
where $d_{\rm z}$ is the luminosity distance and the value of $\Gamma$ is the bulk Lorentz factor estimated with eq. (\ref{gamma}).    Hereafter,  we use $\frac{p-2}{p-1}\simeq 0.29$ with $p=2.4$ the spectral index.

The Klein-Nishina (KN) suppression effect must be considered in SSC emission due to the reduction of the emissivity compared with the classical Compton regime. In the KN regime, the location of the RS SSC cooling break is given by  \citep{2019ApJ...883..162F}
{\small
\be
h\nu^{\rm KN}_{\rm c,r}\simeq 40.9\,{\rm TeV}\, \left(1+Y_{\rm r} \right)^{-1}\,\epsilon^{-1}_{\rm B_r,-1}\,n^{-\frac34}_{-4}\,E^{-\frac14}_{K, 53}\,\Delta^\frac14_{12}\,t^{-\frac12}_1\,,
\ee
}
where $Y_{\rm r}$ is the Compton parameter for the RS due to the KN effect \citep{2010ApJ...712.1232W, 2015MNRAS.454.1073B, 2019ApJ...883..162F}.\\ 

Taking into account the electron Lorentz factors, the spectral breaks and the maximum flux,  the SSC light curves for the fast- and slow-cooling regime are 
{\small
\begin{eqnarray}
\label{fast}
F^{\rm ssc}_{\rm \nu, r}\propto \cases{ 
 t^{\frac{3}{2}}\, \nu_{\rm }^{\frac13},\hspace{1.3cm} \nu<\nu^{\rm ssc}_{\rm c, r}, \cr
 t^{\frac{1}{4}}\, \nu_{\rm }^{-\frac12},\hspace{1.1cm} \nu^{\rm ssc}_{\rm c, r}<\nu<\nu^{\rm ssc}_{\rm m, r}, \cr
 t^{\frac{p}{4}}\,\nu_{\rm }^{-\frac{p}{2}},\,\,\,\,\, \hspace{0.8cm} \nu^{\rm ssc}_{\rm m, r}<\nu\,, \cr
}
\end{eqnarray}
}
and
{\small
\begin{eqnarray}
\label{slow_before}
F^{\rm ssc}_{\rm \nu, r}\propto \cases{ 
t^{\frac{5}{6}}\, \nu_{\rm }^{\frac13},\hspace{1.6cm} \nu<\nu^{\rm ssc}_{\rm m, r}, \cr
t^{\frac{p+3}{4}}\, \nu_{\rm }^{-\frac{p-1}{2}},\hspace{0.7cm} \nu^{\rm ssc}_{\rm m, r}<\nu<\nu^{\rm ssc}_{\rm c, r}, \cr
t^{\frac{p}{4}}\,\nu_{\rm }^{-\frac{p}{2}},\,\,\,\,\,\hspace{1.1cm}\nu^{\rm ssc}_{\rm c, r}<\nu\,, \cr
}
\end{eqnarray}
}
respectively,  where $\nu$ is the observed frequency at a given time $t$.

\paragraph{After the shock crossing time ($t > t_{\rm x} $)}   During this period, the post-shock magnetic field evolves as  $B'\propto t^{-\frac{13}{24}}$  and the Lorentz factor of the lowest-energy electrons, $\gamma_{\rm m,r}\propto t^{-\frac{13}{48}}$.  Electrons above $\gamma_{\rm c,r}\propto \gamma^{-1}_3 B^{-2} t^{-1}\propto   t^{\frac32}$ cool efficiently by synchrotron emission. The corresponding RS synchrotron break frequencies evolve as $\nu^{\rm syn}_{\rm m, r} \propto \,t^{-\frac{73}{48}}$ and $\nu^{\rm syn}_{\rm c, r} \propto t^{\frac{1}{16}}$, while the spectral peak flux density, $F^{\rm syn}_{\rm max, r} \propto   t^{-\frac{47}{48}}$ \citep{2000ApJ...545..807K}.\\

Based on these assessments,  the spectral breaks and the maximum flux of SSC emission are given by

{\small
\bary\label{ssc_after}
h \nu^{\rm ssc}_{\rm m, r}&\simeq&  10.1\,{\rm keV}  \left(1+z\right)^{-1}\,\epsilon^4_{\rm e_r} \epsilon^{\frac12}_{\rm B_r,-4} n^\frac34_{-4}   \Gamma^4_{3}\, \Delta^{\frac{45}{16}}_{12} \,E^{-\frac14}_{\rm K,53}\,t^{-\frac{33}{16}}_{1}\cr
h \nu^{\rm ssc}_{\rm c, r}&\simeq& 101.3\,{\rm TeV}\left(1+z \right)^{3}\,\epsilon^{-\frac72}_{\rm B_r,-4} n^{-\frac94}_{-4} \left(1+Y_{\rm r}\right)^{-4}\Delta^{-\frac{65}{48}}_{12}\cr
&&\hspace{5.7cm} \times\,E^{-\frac{10}{8}}_{\rm K,53} t^{\frac{53}{48}}_{1}\cr
F^{\rm ssc}_{\rm max, r}&\simeq& 2.8\times 10^{-5}\,{\rm mJy}\,\,   \epsilon^{\frac12}_{\rm B_r,-4}\,  \Gamma^{-1}_{3}\, n_{-4}\,  \Delta^{\frac{17}{48}}_{12}\, d^{-2}_{\rm z,28}\,E^{\frac32}_{\rm K,53}\,t^{-\frac {41}{48}}_{1}.\,\,\,\,\,\,\,\,
\eary
}

The break energy above the KN regime is \citep{2010ApJ...712.1232W, 2019ApJ...883..162F}

{\small
\be
h \nu^{\rm KN}_{\rm c,r}\simeq 16.8\,{\rm TeV}\, \left(1+Y_{\rm r} \right)^{-1}\,\epsilon^{-1}_{\rm B_r,-1}\,n^{-\frac34}_{-4}\,E^{-\frac14}_{\rm KN, 53}\,\Delta^{-\frac13}_{12}\,t^{\frac{1}{12}}_{1}\,.
\ee
}

Using the eq. (\ref{ssc_after}) and the evolution of the electron Lorentz factors,  the SSC light curves for the fast- and the slow-cooling regime is given by 

{\small
\begin{eqnarray}
\label{fcssc_t}
F^{\rm ssc}_{\rm \nu, r} \propto \cases{ 
t^{-\frac{11}{9}}\, \nu_{\rm }^{\frac13},\hspace{1.3cm} \nu<\nu^{\rm ssc}_{\rm c, r}, \cr
t^{-\frac{29}{96}}\, \nu_{\rm }^{-\frac12},\hspace{1.1cm} \nu^{\rm ssc}_{\rm c, r}<\nu<\nu^{\rm ssc}_{\rm m, r}, \cr
t^{\frac{70-99p}{96}}\,\nu_{\rm }^{-\frac{p}{2}},\,\,\,\,\hspace{0.55cm}\nu^{\rm ssc}_{\rm m, r}<\nu\,, \cr
}
\end{eqnarray}
}
and
{\small
\begin{eqnarray}
\label{slow_after}
F^{\rm ssc}_{\rm \nu, r} \propto \cases{ 
t^{-\frac{1}{6}}_1\, \nu_{\rm}^{\frac13},\hspace{1.7cm} \nu<\nu^{\rm ssc}_{\rm m, r}, \cr
t^{\frac{17-99p}{96}}\, \nu_{\rm }^{-\frac{p-1}{2}},\hspace{0.7cm} \nu^{\rm ssc}_{\rm m, r}<\nu<\nu^{\rm ssc}_{\rm cut}, \cr
0,\,\,\,\,\hspace{2.3cm}   \nu^{\rm ssc}_{\rm cut}<\nu\,, \cr
}
\end{eqnarray}
}
respectively, where  $\nu^{\rm ssc}_{\rm cut}=\nu^{\rm ssc}_{\rm c, r} \left(\frac{t}{t_{\rm x}}\right)^{-\frac{73}{48}}$. The term $\left(\frac{t}{t_{\rm x}}\right)^{-\frac{73}{48}}$ is introduced due to the adiabatic expansion of the fluid \citep{2000ApJ...545..807K}.  After the peak,  the flux above energy  $h\nu^{\rm ssc}_{\rm cut}$ should disappear. Nonetheless, the angular time delay effect  prevents this disappearance \citep{2000ApJ...543...66P, 2003ApJ...597..455K}.  During this phase,  the evolution of the SSC flux due to high-latitude afterglow emission is described by $F^{\rm ssc}_{\rm \nu, r}\propto t^{-(\beta+2)}$,  where the spectral index\footnote{Hereafter, we use, in general, the notation $F_{\nu}\propto t^{-\alpha}\nu^{-\beta}$ for the closure relations.} can take the values of  $\beta=\frac12$ for $\nu^{\rm ssc}_{\rm c,r}<  \nu<\nu^{\rm ssc}_{\rm m,r}$, $\frac{p-1}{2}$ for $\nu^{\rm ssc}_{\rm m,r}<  \nu<\nu^{\rm ssc}_{\rm c,r}$ and  $\frac{p}{2}$ for $\{\nu^{\rm ssc}_{\rm c,r}, \nu^{\rm ssc}_{\rm m,r}\}< \nu$. It is worth nothing that if off-axis flux dominates, the  SSC light curves follow the high-latitude afterglow emission \citep[e.g., see][]{2007ApJ...655..391K}.

\subsubsection{The thin-shell regime ($\Gamma< \Gamma_{\rm c}$)}
In the thin-shell regime, the RS becomes mildly relativistic.   In this case, the shock crossing time is longer than the duration of the burst ($T_{90}< t_{\rm x}$).  As before, we describe the dynamics before and after the shock crossing time, separately. 
\paragraph{Before the shock crossing time ($t < t_{\rm x}$)}
Given the evolution of the Lorentz factor of the fluid just behind the RS $\gamma_3\propto t^{0}$, the post-shock magnetic field  $B'\propto t^{0}$ and the minimum and cooling electron Lorentz factors  $\gamma_{\rm m, r}\propto t^3$ and $\gamma_{\rm c, r}\propto   t^{-1}$, respectively, the corresponding synchrotron break frequencies at the RS region evolve as $\nu^{\rm syn}_{\rm m, r}\propto t^6$ and $\nu^{\rm syn}_{\rm c, r}\propto t^{-2}$  and the maximum flux in terms of the spectrum as $F^{\rm syn}_{\rm max,r }\propto t^\frac32$ \citep{2000ApJ...545..807K}.\\

For this time interval,   the spectral breaks and the maximum flux of SSC emission are given by

{\small
\bary\label{ssc_before}
h \nu^{\rm ssc}_{\rm m, r}&\simeq& 1.8\times 10^{2}\,{\rm eV}\, \left(1+z \right)^{-1}\,\epsilon^4_{\rm e_r,-1} \epsilon^{\frac12}_{\rm B_r,-1}\,n^{\frac{9}{2}}_{-1} \Gamma^{34}_{2}\, E^{-4}_{\rm K, 51}\,t^{12}_{2}\cr
h \nu^{\rm ssc}_{\rm c, r}&\simeq& 0.3\,{\rm GeV}\,\left(1+z \right)^{3}\,\left(1+Y_{\rm r}\right)^{-4}  \epsilon^{-\frac72}_{\rm B_r,-1}\,n^{-\frac{7}{2}}_{-1}\, \Gamma^{-10}_{2} \, t^{-4}_{2}\,,\cr
F^{\rm ssc}_{\rm max,r} &\simeq& 1.4\times 10^{-4}\,{\rm mJy} \,\epsilon^{\frac12}_{\rm B_r,-1} \, n^{2}_{-1}\, \Gamma^{7}_{2}\, d^{-2}_{\rm z,28}\,E^{\frac12}_{\rm K, 51}\, t^{\frac{5}{2}}_{2}\,,
\eary
}

where the bulk Lorentz factor is less than the critical one ($\Gamma<\Gamma_c\simeq161.9\,E^{\frac18}_{\rm K, 51}\,n^{-\frac18}_{-1}\,T^{-\frac38}_{90}$ for $T_{90}=50\, {\rm s}$). \\ 

The break energy above the KN regime is
{\small
\bary
h \nu^{\rm KN}_{\rm c,r}&\simeq& 111.6\,{\rm GeV}\, \left(1+Y_{\rm r}\right)^{-1}\,\epsilon^{-1}_{\rm B_r,-1}\,n^{-1}_{-1}\,\Gamma^{-2}_{2}\,t^{-1}_{2}\,.
\eary
}

In analogy to the description of the SSC light curves for the thick-shell regime,  the SSC light curves before the shock-crossing time for the fast- and the slow-cooling regime are 

{\small
\begin{eqnarray}
\label{fast_before_thin}
F^{\rm ssc}_{\rm \nu, r} \propto \cases{ 
 t^{\frac{23}{6}}\, \nu_{\rm }^{\frac13},\hspace{1.5cm} \nu<\nu^{\rm ssc}_{\rm c, r}, \cr
 t^{\frac{1}{2}}\, \nu_{\rm }^{-\frac12},\hspace{1.4cm} \nu^{\rm ssc}_{\rm c, r}<\nu<\nu^{\rm ssc}_{\rm m, r}, \cr
 t^{\frac{12p -11}{2}}\,\nu_{\rm }^{-\frac{p}{2}},\,\,\,\,\, \hspace{0.45cm} \nu^{\rm ssc}_{\rm m, r}<\nu\,, \cr
}
\end{eqnarray}
}
and
{\small
\begin{eqnarray}
\label{slow_before_thin}
F^{\rm ssc}_{\rm \nu, r} \propto \cases{ 
t^{-\frac{3}{2}}\, \nu_{\rm }^{\frac13},\hspace{1.6cm} \nu<\nu^{\rm ssc}_{\rm c, r}, \cr
t^{\frac{12p-7}{2}}\, \nu_{\rm }^{-\frac{p-1}{2}},\hspace{0.7cm} \nu^{\rm ssc}_{\rm m, r}<\nu<\nu^{\rm ssc}_{\rm c, r}, \cr
 t^{\frac{12p -11}{2}}\,\nu_{\rm }^{-\frac{p}{2}},\,\,\,\,\,\hspace{0.6cm}\nu^{\rm ssc}_{\rm c, r}<\nu\,. \cr
}
\end{eqnarray}
}

\paragraph{After the shock crossing time ($t > t_{\rm x} $)}
Given the evolution of the the post-shock magnetic field  $B' \propto t^{-\frac47}$ and the minimum and cooling electron Lorentz factors  $\gamma_{\rm m, r}\propto t^{-\frac27}$ and $\gamma_{\rm c,r}\propto  t^{\frac{19}{35}}$, respectively, the corresponding synchrotron break frequencies at the RS region evolve as  $\nu^{\rm syn}_{\rm m, r}\propto t^{-\frac{54}{35}}$ and $\nu^{\rm syn}_{\rm c, r}\propto t^{\frac{4}{35}}$  and the maximum flux in terms of the spectrum as $F^{\rm syn}_{\rm max,r }\propto t^{-\frac{34}{35}}$ \citep{2000ApJ...545..807K}.\\

Following the same previous process,  the spectral breaks and the maximum flux of SSC emission are given by
{\small
\bary\label{ssc_before}
h \nu^{\rm ssc}_{\rm m, r}&\simeq& 1.3\times 10^{-2}\,{\rm eV} \left(1+z\right)^{-1}\,\epsilon^4_{\rm e_r,-1} \epsilon^{\frac12}_{\rm B_r,-1} \,n^{-\frac{43}{210}}_{-1}\, \Gamma^{-\frac{382}{105}}_{2}\, \cr
&&\hspace{5.4cm}\times  \, E^{\frac{74}{105}}_{\rm K,51}\,t^{-\frac{74}{35}}_{2}\,\cr
h \nu^{\rm ssc}_{\rm c, r}&\simeq& 10.2\,{\rm TeV} \left(1+z\right)^{3}\,\left(1+Y_{\rm r}\right)^{-4}  \epsilon^{-\frac72}_{\rm B_r,-1}\,n^{-\frac{53}{30}}_{-1} \, \Gamma^{\frac{58}{15}}_{2}\,\cr
&&\hspace{5.4cm} \times\, E^{-\frac{26}{15}}_{\rm K,51}\, t^{\frac{6}{5}}_{2}\,\cr
F^{\rm ssc}_{\rm max,r} &\simeq&  7.2\times 10^{-5}\,{\rm mJy}\,  \epsilon^{\frac12}_{\rm B_r,-1} \, n^{\frac{191}{210}}_{-1}\, \Gamma^{-\frac{181}{105}}_{2}\, d^{-2}_{\rm z,28}\,E^{\frac{167}{105}}_{\rm K,51}\, t^{-\frac{27}{35}}_{2}\,.\cr
&&\hspace{5.5cm}\eary
}

The break energy above the KN regime is given by \citep{2010ApJ...712.1232W, 2019ApJ...883..162F}

{\small
\bary
h\nu^{\rm KN}_{\rm c, r}&\simeq& 1.9\,{\rm TeV}  \, \left(1+Y_{\rm r}\right)^{-1}\,\epsilon^{-1}_{\rm B_r,-1}\,n^{-\frac{11}{15}}_{-1} \,\Gamma^{\frac{2}{15}}_{2}\,E^{-\frac{4}{5}}_{\rm K, 51}\,t^{-\frac{1}{5}}_{2}.\,\,\,
\eary
}
In analogy to the description of the SSC light curves for the thick-shell regime,  the SSC light curves after the shock-crossing time for the fast- and the slow-cooling regime are
{\small
\begin{eqnarray}
\label{fast_after_thin}
F^{\rm ssc}_{\rm \nu, r} \propto \cases{ 
 t^{-\frac{41}{35}}\, \nu_{\rm }^{\frac13},\hspace{1.4cm} \nu<\nu^{\rm ssc}_{\rm c, r}, \cr
 t^{-\frac{6}{35}}\, \nu_{\rm }^{-\frac12},\hspace{1.2cm} \nu^{\rm ssc}_{\rm c, r}<\nu<\nu^{\rm ssc}_{\rm m, r}, \cr
 t^{\frac{31-37p}{35}}\,\nu_{\rm }^{-\frac{p}{2}},\,\,\,\,\, \hspace{0.5cm} \nu^{\rm ssc}_{\rm m, r}<\nu\,, \cr
}
\end{eqnarray}
}
and
{\small
\begin{eqnarray}
\label{slow_after_thin}
F^{\rm ssc}_{\rm \nu, r} \propto \cases{ 
t^{-\frac{1}{15}}\, \nu_{\rm }^{\frac13},\hspace{1.5cm} \nu<\nu^{\rm ssc}_{\rm m, r}, \cr
t^{\frac{10-37p}{35}}\, \nu_{\rm }^{-\frac{p-1}{2}},\hspace{0.6cm} \nu^{\rm ssc}_{\rm m, r}<\nu <\nu^{\rm ssc}_{\rm cut}, \cr
0,\,\,\,\,\,\hspace{2.1cm}\nu^{\rm ssc}_{\rm cut}<\nu\,. \cr
}
\end{eqnarray}
}

The SSC emission from the RS region could decay faster due to the angular time delay effect, and the corresponding light curve evolution is the same as that discussed at the end of section 2.1.1. \\

\subsection{Synchrotron light curves from FS}\label{syn_LC}
The dynamics of the FS for a relativistic outflow expanding into a uniform-density medium  is explained in  \cite{1998ApJ...497L..17S}.   Using the evolution of synchrotron energy breaks ($\nu^{\rm syn}_{\rm m, f}\propto t^{-\frac32}$ and  $\nu^{\rm syn}_{\rm c, f}\propto t^{-\frac12}$) and the maximum flux ($F^{\rm syn}_{\rm max,f} \propto t^0$), the observed flux in the fast-cooling regime is proportional to {\small $\propto t^{-\frac{1}{4}} \,\nu^{-\frac{1}{2}}$} for {\small $\nu<\nu^{\rm syn}_{\rm m,f}$} and  {\small $\propto  t^{-\frac{3p-2}{4}}\,\nu^{-\frac{p}{2}}$} for {\small $\nu^{\rm syn}_{\rm m,f}<\nu$}.  Hereafter, the sub-index ``f" refers to the derived quantities in the FS.    In the slow-cooling regime, the observed flux is proportional to {\small $\propto t^{-\frac{3p-3}{4}}\,\nu^{-\frac{p-1}{2}}$ for $\nu <\nu^{\rm syn}_{\rm c,f}$} and {\small $\propto t^{-\frac{3p-2}{4}}\,\nu^{-\frac{p}{2}}$} for {\small $\nu^{\rm syn}_{\rm c,f}<\nu$}, where the proportionality factors are explicitly written in e.g., \citet{2016ApJ...831...22F}. \\

\subsection{The GeV flares and the break in the long-lasting emission}

\begin{figure}[h!]
{ \centering
\resizebox*{0.95\textwidth}{0.8\textheight}
{\includegraphics{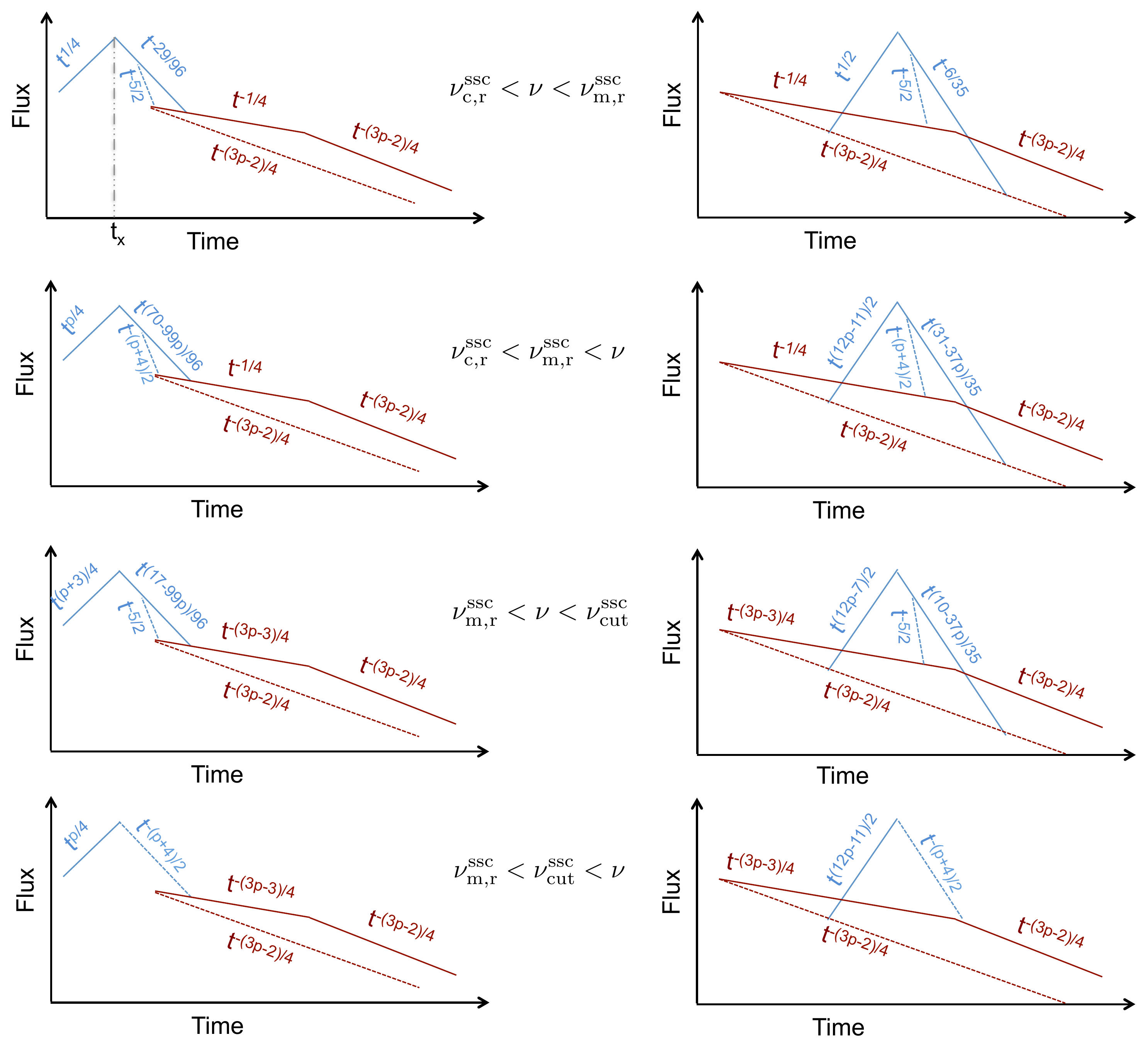}}
}
\caption{Theoretically predicted SSC light curves from RS (blue lines) together with the synchrotron light curves from FSs (red lines) evolving in the fast- and the slow-cooling regime.  The predicted SSC light curves are presented in the thick- (left column) and thin- (right column) shell regime for a uniform-density medium.    The light curves from top to bottom display the SSC flux for $\nu^{\rm ssc}_{\rm c,r}< \nu<\nu^{\rm ssc}_{\rm m,r}$ and $\epsilon^{\rm ssc}_{\rm m,r}< \nu$ (in the fast-cooling regime) followed by the SSC flux for $\nu^{\rm ssc}_{\rm m,r}<\nu<\nu^{\rm ssc}_{\rm cut}$ and $\nu^{\rm ssc}_{\rm cut}< \nu_\gamma$ (in the slow-cooling regime).    The double-dotted dashed line in gray refers to shock crossing time ($t_{\rm x}$). The blue dashed lines indicate an alternative PL evolution of the SSC model (see the SSC light curves in subsection \ref{ssc_lightcurves}).  The breaks exhibited in the solid lines (synchrotron light curves) correspond to the transitions between $\nu<\nu^{\rm syn}_{\rm c, f}$ and $\nu^{\rm syn}_{\rm c, f}<\nu$.  The  red dashed lines indicate that initially, the synchrotron light curves lie in the range $\nu^{\rm syn}_{\rm c, f}<\nu$, thus not breaks are expected.}
 \label{fig1:LC_LAT}
\end{figure}

Figure \ref{fig1:LC_LAT} shows the theoretically predicted SSC and synchrotron light curves from RS and FS evolving in the fast- and the slow-cooling regime, respectively.  The predicted SSC  light curves are presented in the thick- (left column) and thin- (right column) shell regime for a uniform-density medium. The light curves in Figure \ref{fig1:LC_LAT} from top to bottom display the SSC flux for $\nu^{\rm ssc}_{\rm c,r}< \nu<\nu^{\rm ssc}_{\rm m,r}$ and $\nu^{\rm ssc}_{\rm m,r}< \nu$ (in the fast-cooling regime) followed by the SSC flux for $\nu^{\rm ssc}_{\rm m,r}<\nu<\nu^{\rm ssc}_{\rm c,r}$ and $\nu^{\rm ssc}_{\rm c,r}< \nu$ (in the slow-cooling regime).   We do not discuss the effects at the self-absorption regime because its contribution is typically significant only at low energies.  \citep[e.g., see][]{2014ApJ...788...70P}.    Similarly, we do not analyze the  SSC light curves for   $\nu <  {\rm min}\{\nu^{\rm ssc}_{\rm m,r}, \nu^{\rm ssc}_{\rm c,r} \}$ because they are relevant at optical and radio bands \citep{2003ApJ...597..455K} and not  at energies around 100 MeV.  Radiation from high-latitudes received after the shock crossing time may prevent the abrupt disappearance of the RS emission \citep{2000ApJ...545..807K}.    Transitions from fast- to slow-cooling regime and from wind to uniform-density medium have not been considered in these light curves.

When the observed light curves consist of a superposition of SSC from the RS and the synchrotron emission from the FS, as described here, we would naturally expect that the traditional closure relations between the light curve evolution and spectral index would not be satisfied.   Only when the SSC emission is suppressed or has decreased below the FS synchrotron emission, would we expect the closure relation to be satisfied.   It is worth noting that depending on the parameter values synchrotron radiation could dominate over SSC emission.    The SSC light curves from RS indicate that rise ($\alpha_{\rm ris}$) and decay ($\alpha_{\rm dec}$) indices could be expected between  $-\frac{p+3}{4} \lesssim \alpha_{\rm ris} \lesssim -\frac14$  and $\frac{29}{96} \lesssim \alpha_{\rm dec} \lesssim \frac{p+4}{2}$ for a thick-shell regime, and $-\frac{12p-7}{2} \lesssim \alpha_{\rm ris}\lesssim -\frac12$  and  $\frac{6}{35} \lesssim \alpha_{\rm dec} \lesssim \frac{p+4}{2}$ for a thin-shell regime, respectively.  For instance, with a typical value of  the spectral index of $p=2.4$  the temporal rise and decay index for $\nu^{\rm ssc}_{\rm m, r}<\nu <\nu^{\rm ssc}_{\rm c, r}$ is $\alpha_{\rm ris}\simeq -1.35$ and $\alpha_{\rm dec}= 2.50$ for the thick-shell regime, and $\alpha_{\rm ris}\simeq -10.90$ and $\alpha_{\rm dec}=2.50$ for the thin-shell regime, respectively.\\

The FS synchrotron light curves in the LAT band for a uniform-density medium show different behaviours associated with transitions between distinct PL segments  (Figure \ref{fig1:LC_LAT}).  The breaks predicted in the synchrotron light curves correspond to the transitions from  $\nu<\nu^{\rm syn}_{\rm m, f}$ to $\nu^{\rm syn}_{\rm m, f}<\nu$ in the fast-cooling regime, and from $\nu<\nu^{\rm syn}_{\rm c, f}$ to $\nu^{\rm syn}_{\rm c, f}<\nu$ in the slow-cooling regime.  The  synchrotron spectral breaks evolve as $\nu^{\rm syn}_{\rm m, f}\propto t^{-\frac32}$ and $\nu^{\rm syn}_{\rm c, f}\propto t^{-\frac12}$ so that transitions between these PL segments in the fast- and the slow-cooling regime are expected to be associated with changes in the temporal indexes and a steepening in the light curve.  Considering an electron energy index of $p=2.4$, the temporal index, in general, varies from $0.25$ to $1.30$ and from $1.05$ to $1.30$ for the fast and the slow-cooling regimes, respectively \citep{1998ApJ...497L..17S}.\\ 

We argue that a LAT-detected burst that exhibits  a GeV flare and a  break in the long-lasting emission can be interpreted in terms of  external shock emission; the GeV flare as SSC emission from the RS and the break in the long-lasting emission as the transition between PL segments of synchrotron radiation from the FS.   A bright peak from the RS is expected at the RS shock crossing time, $t_{\rm x}$. In the thick-shell regime, $t_{\rm x} < T_{90}$, resulting in the RS SSC peak occurring prior to the onset of the FS emission. Conversely, in the thin shell regime where $t_{\rm x} \gtrsim T_{90}$, the RS SSC peak overlaps with FS emission.

\begin{figure}[h!]
{ \centering
\resizebox*{\textwidth}{0.7\textheight}
{\includegraphics{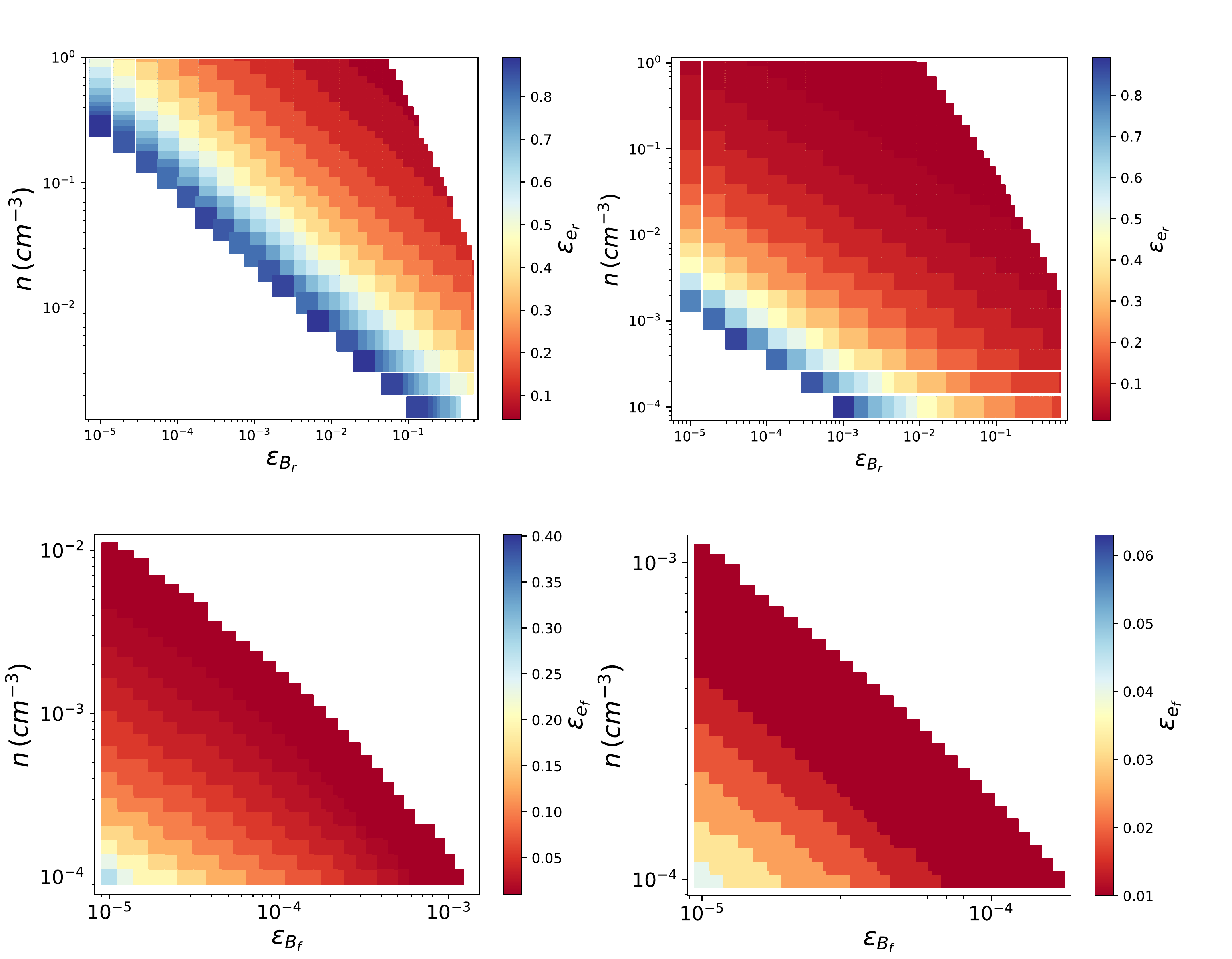}}
}
\caption{Upper panels: Range of microphysical parameters and density for which the SSC emission is above $5\times 10^{-5}\,{\rm mJy}$ for $E=10^{52}\,{\rm erg}$ (left) and  $E=10^{54}\,{\rm erg}$ (right).  The value of the threshold flux was estimated considering the sensitivity of Fermi LAT reported by \cite{2016CRPhy..17..617P, 2017ExA....44...25D}.   Lower panels: Range of microphysical parameters and density for which the FS synchrotron emission exhibits a break  between 100 and 500 s due to the passage of $\nu^{\rm syn}_{\rm c,f}$ for $E=10^{52}\,{\rm erg}$ (left) and  $E=10^{54}\,{\rm erg}$ (right) for p=2.2.}
\label{fig2: parameters}
\end{figure}

As an illustration, we show the parameter space of the microphysical parameters and the density of the circumburst medium for which (i) the RS SSC emission is in the thick-shell regime at 10~s, with a flux density $>5\times10^{-5}$~mJy at 100~MeV (Figure \ref{fig2: parameters}, upper panels), and (ii) the FS cooling frequency, $\nu_{\rm c,f}^{\rm syn}$ crosses 100~MeV between 100 and 500~s after the onset of the burst (Figure \ref{fig2: parameters}, lower panels), resulting in a break in the 100 MeV light curve. We explore two different values of the isotropic-equivalent kinetic energy, $E_{\rm K}=10^{52}$~erg (left column) and $E_{\rm K}=10^{54}$~erg (right column).  The time interval of 100--500~s is chosen to explore the breaks observed in some bursts \citep[e.g. GRB 160509A;][]{Ajello_2019}.   The value of the threshold flux was estimated considering the sensitivity of Fermi-LAT reported by \cite{2016CRPhy..17..617P, 2017ExA....44...25D}.   It is worth noting that the parameter $\epsilon_{\rm e_f}$ is quite strongly constrained from radio peaks in GRB afterglows \citep[e.g. see][]{2017MNRAS.472.3161B}.  

In the model with $E_K = 10^{52}$~erg, there are no values of the physical parameters for which both a GeV flare RS SSC emission and a break in the FS synchrotron radiation due to the passage of $\nu^{\rm syn}_{\rm c,f}$ are simultaneously observed, as long as $R_{\rm e} \equiv \frac{\epsilon_{\rm e_f}}{\epsilon_{\rm e_r}} \approx 1$ and $R_{\rm B} \equiv \left(\frac{\epsilon_{\rm B_r}}{\epsilon_{\rm B_f}} \right)^\frac12 \approx 1$. In the model with $E_{\rm K}=10^{54}$~erg, the two phenomena can be observed in the same burst provided the density is low, $n\approx10^{-3}~{\rm cm}^{-3}$ and if the RS region is highly magnetized, $R_{\rm B} \gg 1$. Indeed, such large magnetizations are expected in magnetically dominated models for the GRB emission \citep[e.g.,][]{2008A&A...480..305G, 2011ApJ...726...90Z, 2012MNRAS.419..573M, 2014MNRAS.445.3892B, 2015ApJ...800...89S, 2016MNRAS.459.3635B, 2017MNRAS.467.2594B, 2017MNRAS.468.3202B, 2018MNRAS.476.1785B}.   Similarly, it can be inferred that the GeV flare in the LAT light curves is unlikely in a weak (e.g. low-luminosity) GRB, even in a low-density environment. Moderately high values of magnetization also have been inferred from several multi-wavelength RS studies, GRB~130427A ($\mathcal{R}_{\rm B}\approx 1$--5; \citealt{lab+13,pcc+14}), GRB~160509A ($\mathcal{R}_{\rm B}\approx3$; \cite{lab+16}), GRB~160625B  ($\mathcal{R}_B\approx1$--10; \cite{alb+17}), although see also \cite{lves+19} for a system with $\mathcal{R}_{\rm B} =0.6\pm0.1$ detected by Fermi-LAT. We thus infer that the presence of simultaneous RS SSC emission and a break in the FS MeV light curve suggests a system with $R_{\rm e} \gtrsim 1$ and/or a combination of high $E_{\rm K}$, a low-density environment, and strong relative magnetization $R_{\rm B}\gtrsim1$ between the RS and FS.

We now apply these principles to investigate the LAT light curve of GRB 160509A. Later on in Section 4, we provide high-energy light curve predictions in the thin-shell regime, using the example of GRB 180418A.

\section{GRB 160509A: The thick-shell case} 

At 08:59:04.36 UTC on 2016 May 09, both instruments onboard Fermi satellite,  GBM (Gamma Burst Monitor), and LAT triggered and located GRB 160509A \citep{2016GCN.19411....1R, 2016GCN.19403....1L}.   The burst was located with coordinates ${\rm R.A.}= 310.1$ and ${\rm Dec}=76.0$   (J2000) with an error radius of $0.50^\circ$ (90\% containment, the systematic error only).  The LAT instrument detected  a very energetic photon with an energy of  52 GeV, at $\approx77$~s after the trigger.  The GBM light curve in the 50 - 300 keV energy range exhibited multiple peaks with a duration of $T_{90}=371$ s.  The Gemini North telescope observed this burst at 13:15 UT on 10 May 2016, obtaining optical spectroscopy and near-IR imaging with GMOS-N and NIRI instruments, respectively. A single and well-defined emission line ([OII] 3727 \AA ) in the spectroscopic analysis indicated a redshift of z=1.17  \citep{2016GCN.19419....1T}.  In the radio bands,  this burst was detected with the VLA  (Very Large Array) at frequencies spanning between 1.3 and 37 GHz, beginning 0.36 days after the trigger time \citep{2016ApJ...833...88L}. \\

\subsection{Fermi-LAT observations}
\begin{figure}[h!]
{ \centering
\resizebox*{0.7\textwidth}{0.35\textheight}
{\includegraphics{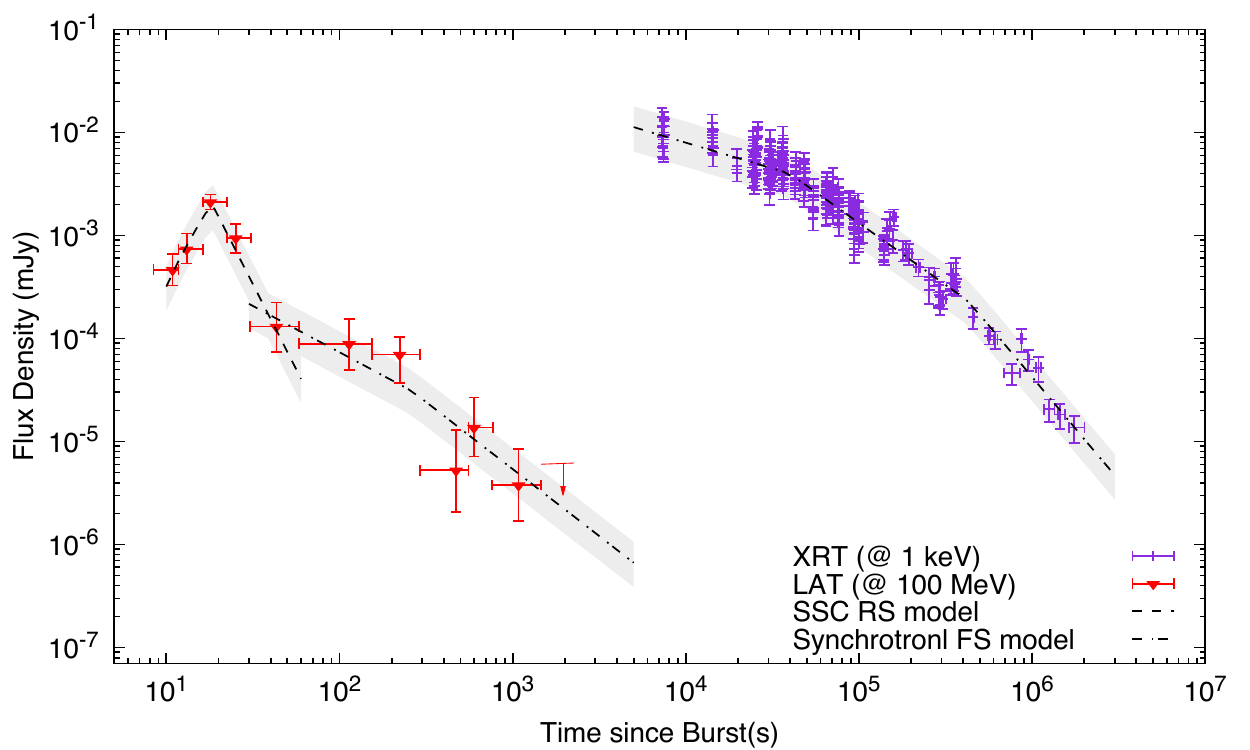}}
}
\caption{Multiwavelength light curves of GRB 160509A, with the synchrotron FS and SSC RS models.  The synchrotron FS  model was used  to describe the long-lived LAT (dashed-dotted line), and  X-ray (dotted line) observations  and the SSC RS model  was used to interpret the GeV flare (dashed line).  The hatched area shows the uncertainty of the best-fit parameters obtained with the MCMC (see Table \ref{table3}).} %Optical R-band data points with the fit were taken from \cite{2016ApJ...833...88L}.
 \label{fig3:multiwavelength}
\end{figure}

%temporally-extended (hereafter called ``long-lasting'') emission
The Fermi-LAT light curve of GRB~160509A\footnote{LAT data points are taken from https://www-glast.stanford.edu/pub\_data/953/.  The energy range of 0.1 - 100 GeV integrated over the spectrum is used to convert from ${\rm erg\, cm^{-2}\,s^{-1}}$ to ${\rm \mu Jy}$ at 100 MeV.} exhibits two components: a GeV flare with a duration of $\sim$ 20 s and a long-lasting emission that begins $\sim$ 40 s after the LAT trigger and extends for more than $\sim 10^3$ s (Figure \ref{fig3:multiwavelength}; red points). We fit the light curve with a series of broken PLs (next two sections) and discuss each component separately.  The best-fit values of the GeV flare and the long-lasting emission are reported in Table \ref{table1}.   It is worth noting that in \cite{Ajello_2019}  the  function defined by \cite{2007ApJ...662.1093W} was used to explore the existence of the plateau phase and to model it simultaneously with the prompt emission component. The best-fit values reported of the PL indices before and after the break time were $0.9\pm0.3$ and $1.3\pm0.3$, respectively \cite[see Table 5 in][]{Ajello_2019}.    Thus, regardless of the function used, the presence of two components is visible.\\

%$\alpha_{\rm L, ris}=1.25\pm0.21$ and $\alpha_{\rm L, dec}=3.14\pm0.65$,
%$\alpha_{\rm L, 1}=0.87\pm0.18$ before $t_{\rm L, br}=316.2\pm61.2\,{\rm s}$ and  $\alpha_{\rm L, 2}=1.26\pm0.21$ after this break time

\subsubsection{The GeV flare}  

To describe the GeV flare, two PL functions were used: $F_{\rm L, pk}\propto (t- t_0)^{-\alpha_{\rm L, ris}}$ for  $t<t_{\rm pk}$ and $\propto t^{-\alpha_{\rm L, dec}}$  for  $t_{\rm pk} < t$.  The term $t_0$ is the starting time of the GeV flare \citep{2007ApJ...655..973K, 2006Natur.442..172V},  $t_{\rm pk}\approx 20\,{\rm s}$ is the time for which the flux reaches the maximum value and begins decreasing, and  $\alpha_{\rm L, ris}$ and $\alpha_{\rm L, dec}$ are the temporal rise and decay indexes, respectively.  Using linear least squares \citep{Lai3034} fitting implemented in ROOT, which is a modular, publicly available scientific software \citep{1997NIMPA.389...81B}, we found that the best-fit values are $t_0=8.4\pm0.5\,{\rm s}$, $\alpha_{\rm L, ris}=-(1.25\pm0.21)$ and $\alpha_{\rm L, dec}=3.14\pm0.65$.

\cite{2011MNRAS.415...77M} investigated a FS origin for the GeV light curves of four GRBs (080916C, 090510, 090902B, and 090926A). They demonstrated that at $t\gg T_{90}$, FS emission could explain the GeV light curve. However, the FS synchrotron contribution usually underpredicts the GeV light curve while the central engine is active (i.e., at $t \lesssim T_{90}$) due to the expectation of on-going energy accumulation in the FS over this period. We note that the GeV flare in GRB~160509A occurs at $t \lesssim T_{90}$. We posit that a possible mechanism to reconcile the discrepancy noted by \cite{2011MNRAS.415...77M} may lie in RS SSC emission.

Comparing the best-fit value $\alpha_{\rm L, ris}=-(1.25\pm0.21)$ with the temporal indexes of the SSC light curves for $t<t_{\rm x}$ (eq. \ref{ssc_before}), we notice that it is consistent with SSC model in the range $\nu^{\rm ssc}_{\rm c, r}<\nu<\nu^{\rm ssc}_{\rm m, r}$ for $p=2.24\pm0.40$. Other PL segments in the SSC light curves would produce atypical values of $p$.  Given the spectral index $p=2.24\pm0.40$ when the SSC emission evolves in the range $\nu^{\rm ssc}_{\rm c, r}<\nu<\nu^{\rm ssc}_{\rm m, r}$, after the shock crossing time the flux decays as $2.13\pm 0.42$  (eq. \ref{ssc_after}) and finally $2+\beta= 2.62\pm0.20$  due to high-latitude emission.  Other PL segments of the SSC light curve  cannot reproduce the decaying phase. Therefore,  the best-fit parameters of the rise and decay PL indexes indicate that the GeV flare  is consistent with a RS evolving in the uniform-density medium and the thick-shell regime.\\ 
The best-fit values of each component are reported in Table \ref{table1}. \\

\subsubsection{The long-lasting emission}

The long-lasting emission component is described with  a broken PL function of the form
\begin{eqnarray}
\label{radio_s1}
F_{\rm L, ee}(t)\propto  \cases{ 
t^{-\alpha_{\rm L, 1}}\,\,\hspace{1.5cm}      t < t_{\rm L, br} , \cr
t^{-\alpha_{\rm L, 2}}\,\, \hspace{1.5cm}      t_{\rm L, br} <t\,.      \cr
}
\end{eqnarray}

The parameters $\alpha_{\rm L, 1}$ and $\alpha_{\rm L, 2}$  are the temporal indexes before and after the time break $t_{\rm L, br}$.  We fit this model to the LAT light curve at $43 - 1081$ using the chi-square method \citep{1997NIMPA.389...81B}. Our best-fit values are $t_{\rm L, br}=316.2\pm61.2\,{\rm s}$,  $\alpha_{\rm L, 1}=0.87\pm0.18$ and $\alpha_{\rm L, 2}=1.26\pm0.28$ which are consistent with those reported in the second LAT GRB catalog \citep[see Fig. 27 in][]{Ajello_2019}.

On the other hand,   \cite{2017ApJ...844L...7T} analyzed the LAT spectrum of the long-lasting emission.   Using a PL function $\frac{dN}{dE}\propto E^{-\Gamma_{\rm L}}$  \cite{2017ApJ...844L...7T} reported  two distinct spectral indexes; $\Gamma_{\rm L,1}=\beta_{\rm L,1}+1=1.42\pm0.12$ for $40 \lesssim t \lesssim 300$ s and  $\Gamma_{\rm L,2}=2.2\pm0.3$ for $300 \lesssim t \lesssim 10^3$ s.   Therefore, this component of the Fermi LAT observations can be described by $F^{\rm syn}_{\rm \nu, f}\propto t^{-0.87\pm0.18} \nu^{-0.42\pm0.12}$ before $\lesssim300$ s and  $F^{\rm syn}_{\rm \nu, f}\propto t^{-1.26\pm0.28} \nu^{-1.2\pm0.3}$ after $\gtrsim300$ s.   In the synchrotron framework (section 2.2), the passage of the synchrotron cooling frequency is expected to result in the steepening of the light curve by $\delta\alpha\approx0.25$, and a change in the spectral index by $\left|\delta\beta\right|\approx0.5$, both of which are consistent with the observed evolution in the LAT light curve and spectrum across $t_{\rm L,br}$. Furthermore, the sign of $\delta\beta$ across the break is a diagnostic of the density profile. The steepening seen here is indicative of a uniform-density environment. This is also consistent with the inference of  \cite{2016ApJ...833...88L}, who infer a uniform-density environment based on modelling the RS emission at radio wavelengths for this burst. Taking $(2-p)/4=1.26\pm0.28$ at $\gtrsim t_{\rm L,br}$ (where the data are less affected by potential contamination from the earlier GeV flare), we find $p=2.35\pm0.37$. This yields $\beta_{\rm L,2}=1.2\pm0.2$, consistent with the observed value. Before the break, we expect a decay rate of $\alpha_{\rm L,1}=3(p-1)/4=1.01\pm0.28$ and spectral slope of $\beta_{\rm L,1}=(p-1)/2=0.68\pm0.19$, consistent with observations. Therefore, the Fermi-LAT observations are consistent with a FS emission in a uniform-density environment, with the spectral ordering $\nu_m < \nu_{\rm LAT} < \nu_{\rm c}$ at $\lesssim316$~s and $\nu_m < \nu_{\rm c}< \nu_{\rm LAT}$ at $\gtrsim316$~s, further implying that $\nu_{\rm c,f}^{\rm syn}$ crosses the LAT band at $\approx316$~s. The value of $p=2.35\pm0.37$ for the FS is also consistent with that previously obtained by \cite{2016ApJ...833...88L} using the X-ray observations for this burst.

\subsection{Constraint on the parameters from the afterglow observations}

\subsubsection{Afterglow observations}

The Swift XRT (X-ray Telescope) followed-up GRB 160509A  in a series of observations (seven) tiled on the sky \citep{2016GCN.19408....1K}.  The data in all modes began to be collected from  $7.3\times 10^3$ s to 20 days after the trigger time.   The XRT data used in this analysis is publicly available in the website database.\footnote{http://www.swift.ac.uk/xrtproducts/}  The XRT flux  was extrapolated from 10  keV to 1 keV with the conversion factor shown in \cite{2010A&A...519A.102E}. In the PC mode, the spectral value of the photon index was $\Gamma_{\rm X}=\beta_{\rm X}+1=1.97^{+0.08}_{-0.07} $ for a galactic (intrinsic) absorption of  $N_{H}= 2.12 (2.7^{+0.04}_{-0.07})\times 10^{22}\,{\rm cm^{-2}}$.\\
\cite{2016ApJ...833...88L} analyzed the entire XRT data using the HEASOFT (v6.18) to fit the spectra.  These authors initially using a broken PL and reported two spectral indexes  of  $\Gamma_{X}=\beta_{X}+1=2.01\pm 0.05$ and $\Gamma_{X}=\beta_{X}+1=2.12\pm 0.05$ for the intervals  $7.3\times 10^3$  to $3.7\times 10^4\, {\rm s}$ and  $4.3\times 10^4$  to $1.3\times 10^6\,{\rm s}$, respectively. Supposing that the spectral index did not evolve during the whole interval, they used a PL and  reported an X-ray spectral index of $\Gamma_{\rm X}=2.07\pm 0.04$ for the entire interval. It is worth noting that the values of the spectral index reported in  the website database and by \cite{2016ApJ...833...88L} are consistent.\\

Magenta data points in Figure \ref{fig3:multiwavelength} show the XRT light curve obtained at 1 keV.\footnote{https://www.swift.ac.uk/burst\_analyser/00020607/}  In  accordance with the shape of  the X-ray light curve, it  is divided in three intervals, labelled II ($3.5\times 10^3 \lesssim t  \lesssim 5.1\times 10^4\,{\rm s}$), III ($5.1\times10^4 \lesssim t \lesssim 4.28\times10^5\,{\rm s}$), and IV ($4.28\times10^5 \,{\rm s} \lesssim  t$).   We model each segment of the light curve with a PL function ($t^{-\alpha_{\rm X}}$) using the chi-square minimization method. The best-fit indexes are $\alpha_{X,II}=  0.50\pm0.12$, $\alpha_{X,III}= 1.23\pm0.13 $ and $\alpha_{X,IV}= 2.06\pm0.3 $, for the intervals II,  III  and IV,   respectively. The best-fit value of each interval with its corresponding chi-square test statistic is reported in Table \ref{table2}.\\ 

Using the closure relations, the flux during the interval III is described by $F^{\rm syn}_{\rm \nu, f}\propto t^{-1.23\pm0.13} \nu^{-1.07\pm0.04}$, which can be understood as FS synchrotron emission  in the regime,  $\nu^{\rm syn}_{\rm c,f}< \nu$ with $p=2.30\pm0.15$.    Given that the spectral index remains unchanged during the intervals II and IV, the interval II with the index $\alpha_{X,II}=  0.50\pm0.12$ is consistent with the plateau phase, while the steepening of the light curve in interval IV ($\alpha_{\rm X,IV}=2.06\pm0.3$) is consistent with post-jet break evolution. 

Our observation that $\nu_{\rm c,f}^{\rm syn}<\nu_{\rm X}$ at $\gtrsim5.1\times10^4\,{\rm s}$, together with our prior inference that $\nu_{\rm c,f}^{\rm syn}\approx100\,{\rm MeV}$ at $\approx300\,{\rm s}$ (Section 3.1.2), implies a rapid evolution of the FS synchrotron cooling frequency\footnote{In fact, $\alpha_{L,1}$ and $\alpha_{\rm X,II}$ are consistent within the error bars, which would suggest that the break in the X-ray light curve at $5.1\times10^4$~s is due to the passage of $\nu_{\rm c,FS}^{\rm syn}$. We verify this in section 3.3.1.}, at least as fast as $\nu_{\rm c,f}^{\rm syn}\propto t^{-2.2}$. This is at variance with the expectation of $\nu_{\rm c,f}^{\rm syn}\propto t^{-1/2}$ in a uniform-density environment \citep{1998ApJ...497L..17S}. Similar rapid evolution of $\nu_{\rm c,f}^{\rm syn}$ has been inferred in other cases, with potential explanations involving time evolution of the microphysical parameters, steep circumburst density profiles, energy injection into the FS, and the proximity of the jet break \citep{2008Natur.455..183R,2011A&A...535A..57F, 2012ApJ...747L..30V}. 

Here, we consider the evolution of microphysical parameters ($\epsilon_{\rm e_f}\propto t^{\alpha_{\rm e}}$, $\epsilon_{\rm B_f}\propto t^{\alpha_{\rm B}}$) as a possible explanation.   The total energy transferred from protons to electrons and magnetic field is not well understood during the FS, so the fraction of electron and the magnetic density during the afterglow could vary  \citep[e.g., see][]{2003ApJ...597..459Y, 2003MNRAS.346..905K, 2006A&A...458....7I,2006MNRAS.369..197F, 2006MNRAS.369.2059P}. We emphasize that this is one possible model with the potential to explain the rapid evolution of $\nu_{\rm c,f}^{\rm syn}$. In this model, the synchrotron light curve in the slow cooling regime is given by

\begin{eqnarray}
\label{evol_par}
F_{\rm \nu}\propto  \cases{ 
t^{\frac{(4\alpha_e-3)(p-1)+\alpha_B(p+1)}{4}}\,\nu^{-\frac{p-1}{2}} \,\hspace{0.5cm}      \nu^{\rm syn}_{\rm m, f} <\nu < \nu^{\rm syn}_{\rm c, f},\,\,\,\,\,\,\, \cr
t^{\frac{-(3p-2)+\alpha_B(p-2) +4\alpha_e(p-1)}{4}} \,\nu^{-\frac{p}{2}}\, \hspace{0.3cm}    \nu^{\rm syn}_{\rm c, f} <\nu \,,      \cr
}
\end{eqnarray}

where the synchrotron spectral breaks evolve as  $\nu^{\rm syn}_{\rm m, f}\propto t^{\frac{-3+4\alpha_e+\alpha_B}{2}}$ and  $\nu^{\rm syn}_{\rm c, f}\propto t^{\frac{-(3\alpha_B +1)}{2}}$.   It is worth noting that once the microphysical parameters stop evolving (i.e., $\alpha_e=0$ and $\alpha_B=0$), the standard synchrotron FS model is recovered  \citep{1998ApJ...497L..17S}.

To summarize, the LAT light curve at $\le40$~s is consistent with SSC emission from a RS in the thick-shell regime and a uniform-density circumburst environment. The LAT data at $>40$~s and the X-ray light curve are consistent with synchrotron emission from the FS in a uniform-density circumburst environment with $p\approx2.4\pm 0.1$  and the spectral ordering $\nu_{\rm m,f} < \nu_{X} < \nu_{\rm c,f}< \nu_{\rm LAT}$ for $300 \lesssim t\lesssim 5.1\times 10^4\,{\rm s}$ and $\nu_{\rm m,f} < \nu_{\rm c,f} < \nu_{X}$ for $t\gtrsim 4.5\times 10^4\,{\rm s}$.

\subsubsection{Constraint on the parameters}

We normalize the synchrotron emission  at $h\nu=$ 100 MeV and 1 keV  for the LAT and  X-ray observations, respectively.   The synchrotron light curves that evolve  in a uniform-density medium before \citep{1998ApJ...497L..17S} and after \citep{1999ApJ...519L..17S} the jet break  were used to describe the long-lasting LAT and X-ray emissions, and the GeV flare observed before $\leq$ 40 s are fitted with the SSC emission in the thick-shell regime (eqs. \ref{slow_before} and \ref{slow_after}).   The synchrotron FS model with varying the microphysical parameters  is  used to describe the X-ray flux during the time interval $10^3$ -  $5.1\times10^{4}$ s (eq. \ref{evol_par}).   The luminosity distance was estimated using the cosmological parameters reported  in \cite{2018arXiv180706209P}.   Using our estimate of $E_{\rm K}=6.977^{+0.801}_{-0.802}\times10^{53}\,{\rm erg}$ and the isotropic-equivalent $\gamma$-ray energy, $E_{\gamma,\rm iso}=(10\pm1)\times10^{52}$~erg \citep{Ajello_2019}, we obtain a prompt gamma-ray efficiency of $\eta=0.15^{+0.02}_{-0.02}$.

\begin{figure}[h!]
	{ \centering
		\resizebox*{\textwidth}{0.65\textheight}
		{\includegraphics[angle=-90]{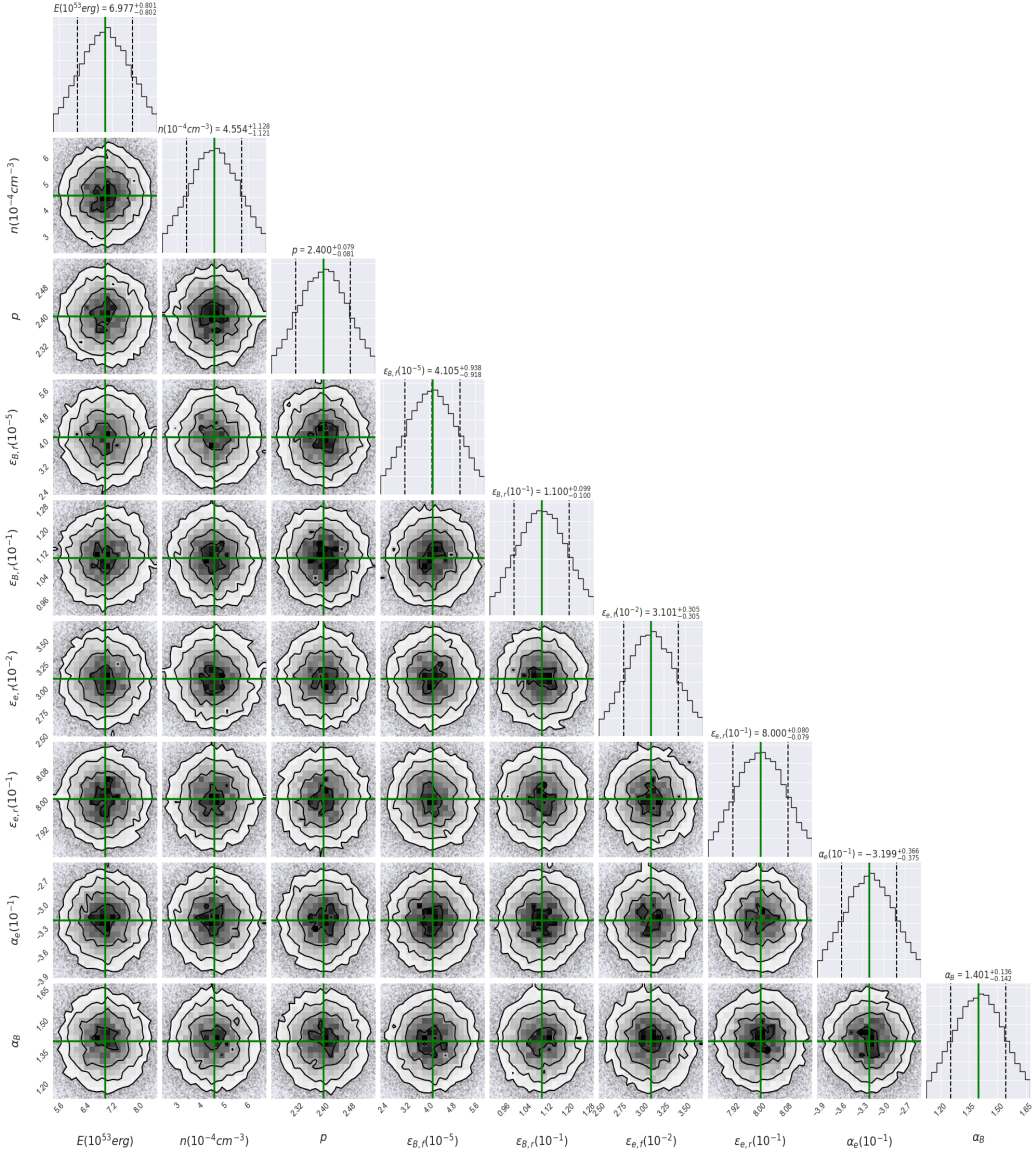}}
	}
	\caption{Corner plot of the parameters derived from fitting the LAT and X-ray light curves of GRB~160509A with a RS-SSC and FS synchrotron model, together with median values (green lines) and 1$\sigma$ credible intervals (dashed lines). MCMC summary statistics for all parameters are listed in Table \ref{table3}.   A set of normal distributions are made for the priors. The values of the mean and standard deviation for each of these normal distributions used for the priors are: $E_{\rm K}(10^{53}\,{\rm erg})=[8.0; 1.0]$,   $n\,(10^{-4}{\rm cm^{-3}})=[10.0;1.0]$,   $p=[2.2;0.1]$,  $\epsilon_{\rm B_f}(10^{-5})=[1.0;0.1]$,  $\epsilon_{\rm B_r}(10^{-1})=[1.0;0.1]$, $\epsilon_{\rm e_f}(10^{-2})=[10.0;1.0]$, $\epsilon_{\rm e_r}(10^{-1})=[1.0; 0.1]$, $\alpha_e(10^{-1})=[-3.0; 0.1]$ and $\alpha_B=[1.0;0.1]$. }
	\label{fig:param_LAT}
\end{figure}

Our analytical afterglow model, as described in Section \ref{sec2}, is completely determined by a set of nine parameters {\small $\Sigma_\sigma=\{E_{\rm K},\, n, \, p,\, \epsilon_{\rm B_f},\, \epsilon_{\rm B_r},\, \epsilon_{\rm e_f},\,  \epsilon_{\rm e_r},\, \alpha_e,\, \alpha_B \}$}. We, then, assign prior distributions to these parameters for application in a Markov-Chan Monte Carlo (MCMC) simulation. Shape-wise we choose Normal distributions for each physical parameter of the system; this allows us to pass the minimum amount of information (and consequently bias) necessary for the simulation. After determining the shape, we then assign a mean and standard deviation for each parameter.  The choice of values is such that our prior distributions cover a range of the typical values found for these parameters in the literature of GRB afterglow modelling \citep[e.g, see][]{2009MNRAS.400L..75K, 2015PhR...561....1K,  Ajello_2019}, while maintaining a reasonable computational time. We then assign a likelihood function described by a Normal distribution whose mean is our afterglow model and standard deviation is a hyperparameter $\sigma$. For this hyperparameter we chose a value that returns a distribution sufficiently large so that the likelihood can explore the region around the detections containing the data uncertainty, similarly to the models we used in \cite{2019ApJ...871..200F}.   We opted to use a Half-Normal distribution, with static standard deviation, to describe this $\sigma$ parameter, this way our likelihood is allowed to better explore the space around the observed data giving more leeway to the sampler. We use the No-U-TurnSampler from the PyMC3 python distribution to generate a total of 17300 samples, allowing a total of 7000 tuning iterations. The results of the MCMC analysis are summarized in Table \ref{table3}.  The best-fitting curves are shown in Figure \ref{fig3:multiwavelength}.  The corner plot displaying the one-dimensional marginalized posterior distributions for each parameter and the two-dimensional marginalized posterior distributions for each pair of parameters is shown in Figure \ref{fig:param_LAT}.  It is worth nothing that a Bayesian technique of checking the ability of the model to explain the observed data are posterior predictive checks \citep{04f33f2e6d5e4ef19974afc287f496db}.   Concerns about the convergence of affine invariant ensemble sampler in high dimension are described by \cite{2015arXiv150902230H}.  We use the Gelman and Rubin's convergence diagnostic \citep{1992StaSc...7..457G} through the parameter $\hat{R}$  for each variable verify the convergence of the sampler. For all variables the $\hat{R}$ returned values $\approx 1$, indicating a well-behaved, converging sampler.

%\citep{2016ascl.soft10016S}

Based on the values reported in Table \ref{table3}, the implications of the results are discussed in the following section.

\subsection{Implications of the Results and Discussion}

\subsubsection{Microphysical parameters}

Given the microphysical parameter  associated with the magnetic field in the RS region ($\epsilon_{\rm B_r}=1.100^{+0.099}_{-0.100}\times 10^{-1}$), the magnetization parameter \cite[see Fig. 6 in][]{2005ApJ...628..315Z} lies in the range $0.04\lesssim\sigma\lesssim 0.2$ which corresponds to a regime for which the RS is produced.  In the RS region, the self-absorption, the characteristic, and the cutoff frequency breaks of synchrotron radiation at 30 s are $3.5\times10^{5}\,{\rm Hz}$, $5.3\times 10^{14}\,{\rm Hz}$ and $1.7\times 10^{16}\,{\rm Hz}$, respectively.  It shows that the synchrotron emission  evolves in the slow-cooling regime and  lies in the weak self-absorption regime so that a  thermal component is not expected by this mechanism \citep{2003ApJ...597..455K}.

The best-fit values of the magnetic field parameters from the FS and RS regions are different.  The evolution of the RS requires that the outflow is moderately magnetized $\mathcal{R_{\rm B}}\approx 35$.\footnote{\cite{2016ApJ...833...88L} reported a value of $\mathcal{R_{\rm B}}\approx 3$.}  

In the radio bands,   \cite{2016ApJ...833...88L} presented a multi-frequency radio detection with the VLA beginning 0.36 days after the trigger time. The VLA observations were carried out at frequencies spanning 1.3 and 37 GHz. These authors found that the X-ray and radio emission originated from distinct regions.  They modelled the radio to X-ray emission as a combination of synchrotron radiation from both the RS and FS.   They showed that the radio observations were dominated up to 10.03 days by synchrotron from the RS region.   Our best-fit values using LAT and X-ray observations, $E_{\rm K}=(6.977^{+0.801}_{-0.802})\times 10^{53}\,{\rm erg}$, $n= (4.554^{+1.128}_{-1.121}) \times 10^{-4}\,{\rm cm^{-3}}$, and $p=2.400^{+0.079}_{-0.081}$ are consistent with the values of \cite{2016ApJ...833...88L}, who found $E_{\rm K}=18.7^{+5.4}_{-2.6}\times 10^{52}\,{\rm erg}$,  $n=(8.6\pm2.2)\times10^{-4}$, and $p=2.39\pm0.03$.  Our values of the FS microphysical parameters ($\epsilon_{\rm B_f}=4.105^{+0.938}_{-0.918} \times 10^{-5}$  and $\epsilon_{\rm e_f}=3.101^{+0.305}_{-0.305} \times 10^{-2}$) are much smaller than reported by \cite{2016ApJ...833...88L} ($\epsilon_{\rm B_f}=0.11^{+0.07}_{-0.05}$ and $\epsilon_{\rm e_f}=0.84^{+0.06}_{-0.08}$); however, we note that both sets of analyses suffer from some degeneracy, since (i) the observations do not allow us to locate the FS synchrotron self-absorption break, and (ii) $\nu_{\rm m,FS}^{\rm syn}$ is not uniquely constrained due to the large optical extinction along the line of sight. Thus, it is possible that the true values of these parameters are intermediate between those derived here and in \cite{2016ApJ...833...88L}.\\

During the plateau phase ($10^3 - 5.1\times 10^4$ s) the microphysical parameters are given by $\epsilon_{\rm e_f}  \left(\frac{t}{t_{\rm k}}  \right)^{\rm \alpha_e}$ and $\epsilon_{\rm B_f}  \left(\frac{t}{t_{\rm k}}  \right)^{\rm \alpha_B}$ with the best-fit values of  $\alpha_e=-3.199^{+0.366}_{-0.375} \times 10^{-1}$ and $\alpha_B= 1.401^{+0.136}_{-0.142}$ for a normalization time fixed to $t_{\rm k}=10^3\,{\rm s}$.  While the magnetic parameter increases with time, the electron density parameter decreases. It is worth noting that  different authors have considered distinct possibilities in the evolution of the microphysical parameters \citep[e.g., see][]{2006MNRAS.369..197F, 2006MNRAS.369.2059P, 2003MNRAS.346..905K, 2006A&A...458....7I}.   With the best-fit values,  the synchrotron break frequencies evolves as $\nu^{\rm syn}_{\rm m, f}\propto t^{-1.44\pm0.05}$ and $\nu^{\rm syn}_{\rm c, f}\propto t^{-2.60\pm0.21}$, and the flux as $t^{-0.31\pm 0.08}$ for $\nu^{\rm syn}_{\rm m, f} <\nu < \nu^{\rm syn}_{\rm c, f}$.  One can see that the best-fit value of the temporal index in the interval II (the plateau phase) reported in Table \ref{table2} agrees with the predicted value of $-0.31\pm 0.08$ derived when the  microphysical parameters evolve with time.  Similarly, the rapid evolution of $\nu^{\rm syn}_{\rm c, f}\propto t^{-2.60\pm0.21}$ indicates that   the breaks observed in the LAT and the X-ray light curve at $\sim 316$ s and $5.1\times 10^4\,{\rm s}$, respectively,  can be  interpreted as the passage of the synchrotron cooling break through the Fermi-LAT  and Swift-XRT bands at 100 MeV and 1 keV, respectively.\\

\subsubsection{The bulk Lorentz factors}

Our analysis of the multi-wavelength afterglow observations  leads to an initial bulk and critical Lorentz factors of  $\Gamma\simeq$ 600 and  $\Gamma_{\rm c}\simeq350$, respectively.\footnote{The bulk Lorentz factor is obtained using eq. \ref{gamma} and the best-fit values reported in Table \ref{table3}. As expected the bulk Lorentz factor is above the critical one.}   This shows our RS model  evolving in the thick-shell regime is self-consistent.\\ 

The break observed in the X-ray observations  at $t_j\simeq 4.5\times 10^5$ s is associated  with a jet break.  This value  leads to a jet opening angle of $\theta_j\simeq 8.3^\circ$ and a bulk Lorentz factor at the jet-break time of $\Gamma_{\rm j, br}$ = 6.9 \citep{1999ApJ...519L..17S}.  The jet opening angle obtained is twice the value reported in \cite{2016ApJ...833...88L}, so their result is not based solely on the X-ray light curve\\

The value of the initial bulk Lorentz factor inferred during the deceleration phase is  similar to those reported by other burst detected by Fermi-LAT \citep{2012ApJ...755...12V}.  Since GRB 160509A exhibited one of the most energetic photons reported by Fermi-LAT in the second GRB catalog, it is expected that the value of the bulk Lorentz factor lies in the range of the brightest LAT-detected bursts \citep[$500 \lesssim \Gamma \lesssim 1000$;][]{2011ApJ...729..114A,  2013ApJ...763...71A, 2009ApJ...706L.138A, 2010ApJ...716.1178A, 2014Sci...343...42A, 2019ApJ...879L..26F, 2019ApJ...885...29F}, as found in this work.\\

\vspace{1cm}

\subsubsection{The ambient density profile}
If we consider the core-collapse scenario for long GRBs \citep[e.g.][]{1993ApJ...405..273W, 1999ApJ...524..262M, 2006ARA&A..44..507W} then stellar winds from the massive star are expected to form the circumburst medium which scale as a function of the radius \cite[e.g.][]{1999ApJ...520L..29C}.  Beyond the wind termination shock, the density profile is expected to transition from a wind-like to the uniform-density interstellar medium \citep{2017ApJ...848...15F}.

\cite{2019ApJ...883..134T} analyzed 26 long bright GRBs and showed that a subset of these events (22 GRBs) could be explained with the uniform-density medium. Other observational studies reached similar conclusions \citep[e.g.,][]{2003ApJ...597..459Y, 2011A&A...526A..23S}.   These outcomes may imply that a wind profile cannot be explained when  the radii of the FS is reached \citep{2011A&A...526A..23S}.

The best-fit value of the circumburst density $n=4.554^{+1.128}_{-1.121}\times 10^{-4}\,{\rm cm^{-3}}$ indicates that GRB 160509A exploded in an environment with very low density comparable to a halo or intergalactic medium with $n\sim 10^{-5} - 10^{-3}\,{\rm cm^{-3}}$.   The inferred low density agrees with our prediction (Section 2.3) that the simultaneous occurrence of a GeV flare and a late-time break in the LAT light curve (due to the passage of $\nu_c^{\rm syn}$) requires $n\lesssim10^{-3}$~cm$^{-3}$ for typical parameters.  A large fraction (22)  of LAT-detected GRBs have been shown to have exploded in an environment best described as a uniform-density medium \citep{2019ApJ...883..134T}, and GRB~160509A continues this trend.

Although \cite{2016ApJ...833...88L} favored a uniform-density environment for this burst on physical grounds (based on the inferred initial Lorentz factor from the RS emission), they could not conclusively distinguish between a wind a uniform-density medium based on X-ray, optical, and radio afterglow observations alone.  As shown in this paper,  the analysis of the LAT observations suggests that this is consistent with the evolution of the external shocks in a uniform-density medium.

\subsubsection{VHE photons above synchrotron limit and SSC emission from FS}

\cite{Ajello_2019} presented in the second Fermi-LAT GRB catalog the bursts with photon energies above $>$ 10 GeV. Two such  photons were associated with GRB 160509A, the first photon with an energy of 51.9 GeV arriving at 76.5 s and the second one of 41.5 GeV arriving at 242 s after the trigger time.   Given the best-fit parameters (see Table \ref{table3}), the maximum energy photons produced by synchrotron radiation during the evolution of the FS are 6.82 and 4.43 GeV at 76.5 and 242 s, respectively.\footnote{We use the upper limit on the energy of photons that can be produced by synchrotron radiation in FSs \citep[e.g., see][]{2009MNRAS.400L..75K, 2019ApJ...883..162F}.}     Furthermore,  RS SSC emission cannot explain the highest-energy photons detected at 76.5 s and 242 s after the trigger time because this component is subdominant to the synchrotron FS emission at $\geq 40$ s.   Therefore, the observed high-energy photons require a process distinct from both FS synchrotron and RS SSC emission.  We now consider whether SSC emission from the FS could explain these photons \citep[e.g., see][]{2015MNRAS.454.1073B, 2019ApJ...883..162F}.\\

Following \cite{2019ApJ...883..162F} and the parameters reported for this burst (see Table  \ref{table3}),  the spectral breaks  and the maximum flux for SSC emission in the FS can be expressed as

{\small
\bary\label{ssc_br-h}
h\nu^{\rm ssc}_{\rm m,f}&\simeq& 7.9\,{\rm GeV}\,  \left(1+z \right)^{\frac54}\,\epsilon_{\rm e_f,-2}^{4}\,\epsilon_{\rm B_f,-5}^{\frac12}\,n^{-\frac14}_{-4}\,E^{\frac34}_{\rm K,53}\,t^{-\frac94}_2,\cr
h\nu^{\rm ssc}_{\rm c,f}&\simeq& 2.9\times 10^{11}{\rm TeV} \left(1+z\right)^{-\frac34}\left(1+Y_{\rm f} \right)^{-4}\epsilon_{\rm B_f,-5}^{-\frac72},\cr
&&\hspace{4.8cm}\times\,n^{-\frac94}_{-4}\,E^{-\frac53}_{\rm K,53}\,t^{-\frac14}_2\,,\cr
F^{\rm ssc}_{\rm max, f}&\simeq& 2.8\times 10^{-8}\,{\rm mJy}\left(1+z\right)^{\frac34}\,\epsilon_{\rm B_f,-5}^{\frac12}\,n^{\frac54}_{-4}\,d^{-2}_{\rm z, 28}\,E^{\frac54}_{\rm K,53}\, \cr
&&\hspace{6.cm}\times\,t^{\frac14}_2,\,\,\,\,\,\,\,\,
\eary
}

where $Y_{\rm f}$ is the Compton parameter for the FS \citep{2010ApJ...712.1232W, 2015MNRAS.454.1073B, 2019ApJ...883..162F}.    In the slow-cooling regime the SSC light curve is given by \cite[e.g][]{2019ApJ...883..162F}

{\footnotesize
\begin{eqnarray}
\label{ssc_ism}
F^{\rm ssc}_{\nu,f}=  \cases{ 
A_{\rm l}\,t_2\,  (h \nu)_{10}^{\frac13},\hspace{2.2cm} \nu <\nu^{\rm ssc}_{\rm m,f}, \cr
A_{\rm m}t^{-1.33}_2 (h \nu)_{10}^{-0.7}, \hspace{1.3cm} \nu^{\rm ssc}_{\rm m,f}<\nu <\nu^{\rm ssc}_{\rm c,f},\hspace{.25cm}\cr
A_{\rm h}t^{-1.45}_2\,(h \nu)_{10}^{-1.2},\,\,\,\, \hspace{1.1cm}  \nu^{\rm ssc}_{\rm c,f} <\nu\,, \cr
}
\end{eqnarray}
}
  
where $(h \nu)_{10}=10\,{\rm GeV}$ and $t_2=100\,{s}$ correspond to the energy band and timescale of this process,  and the coefficients are given by

{\footnotesize
\bary
A_{\rm l}&\simeq& 1.9\times 10^{-8}\,{\rm mJy}\,\left(1+z \right)^\frac13\,\epsilon_{\rm B_f,-5}^{\frac13}\,\epsilon_{e_f,-2}^{-\frac{4}{3}}\,n^\frac43_{-4}\, d^{-2}_{\rm z, 28}\,E_{\rm K,53}\,,\cr
A_{\rm m}&\simeq&2.6\times 10^{-8}\,{\rm mJy}  \left(1+z \right)^{1.63}\epsilon_{\rm B_f,-5}^{0.85}\,\epsilon_{e_f,-2}^{2.8}\,n^{1.08}_{-4} \,d^{-2}_{\rm z, 28}\,\cr
&&\hspace{6.2cm}\times E^{1.78}_{\rm K,53}\,,\cr
A_{\rm h} &\simeq& 9.7\times 10^{-2}\,{\rm mJy} \left(1+z \right)^{1.25}\,\left(1+Y_{\rm f}\right)^{-2}\, \epsilon_{\rm B_f,-5}^{-0.45}\,\epsilon_{e_f,-2}^{2.8}\,\cr
&&\hspace{5cm}\times\,\,d^{-2}_{\rm z, 28}\,E^{1.15}_{\rm K,53}\,. 
\eary
}

Given eqs. (\ref{ssc_br-h}) and (\ref{ssc_ism}),  the two energetic photons at 76.5 s and 242 s associated with GRB 160509A lie in the range of  $\nu^{\rm ssc}_{\rm m,f}<\nu <\nu^{\rm ssc}_{\rm c,f}$. In this case, the number of VHE photons ($N_{\gamma}$) arriving to LAT effective area ($A$) at 100 s can be estimated as $N_{\gamma}\sim 1\,{\rm ph} \left(\frac{F^{\rm ssc}_{\nu,f}}{10^{-8}\,{\rm \frac{erg}{cm^{2}\,s}}}\right) \left(\frac{100\,{\rm s}}{t} \right)\left(\frac{10^4\,{\rm cm^{2}}}{A}\right) \left(\frac{h\nu}{10\,{\rm GeV}}\right)$ which is consistent with observations. The conversion from $1\,{\rm mJy}$  normalized at $10\,{\rm GeV}$ to $\simeq 0.2 \,{\rm erg\,cm^{-2}\,s^{-1}}$ in the (0.1 - 100) GeV energy range for an spectral index of 2 is used.

Taking into account the distance to GRB 160509A,  the SSC flux must be corrected by the extragalactic background light (EBL) attenuation.  Using the model proposed in \cite{2017A&A...603A..34F}, the attenuation factor $\exp\{-\tau(\epsilon_\gamma, z) \}$ at $h\nu=100\, {\rm GeV}$ and $z=1.17$ is  $\approx$ 0.5.

The frequency above which KN effects are important can be written as
{
\small
\bary
h \nu^{\rm KN}_{\rm c,f}&\simeq& 3.9\times10^3\,{\rm TeV}\, \left(1+z \right)^{-\frac34}\, \left(1+Y_{\rm f}\right)^{-1} \,\epsilon_{\rm B_f,-4}^{-1}\,n^{-\frac34}_{-4}\cr
&&\hspace{4.5cm}\times\, \,E^{-\frac14}_{\rm K,53}\,t^{-\frac14}_2\,,
\eary
}

which is above the energy range considered.   Given the minimum ($\gamma_{\rm m}$) and cooling ($\gamma_{\rm c}$) electron Lorentz factors shown in \cite{2019ApJ...883..162F}, the synchrotron and SSC luminosity ratio can be computed as \citep{2001ApJ...548..787S}

\be
\frac{L^{\rm ssc}_\nu}{L^{\rm syn}_\nu}\simeq 0.7\, n_{-4}\, R_{18}\,\gamma_{\rm c,6}^2\left(\frac{\gamma_{\rm c,7}}{\gamma_{\rm m,3}}\right)^{-1.4}\,,
\ee

where $R=9.9\times10^{17}\,{\rm cm} \left(1+z \right)^{-\frac14}\, n^{-\frac14}_{-4}\,E^{\frac14}_{\rm K,53}\,t^{\frac14}_2$ is the FS radius.   In the case of a uniform medium, 70\% of the synchrotron luminosity is up-scattered by SSC emission.   The KN suppression for the SSC photon is important above $\sim 10^3\,{\rm TeV}$.

The HAWC observatory performed a search for VHE (0.1 - 100 TeV) photons in temporal (using four search windows) and spatial coincidence with GRB 160509A \citep{2016GCN.19423....1L}.  One of these time windows corresponds to around the time (from - 20  to 20 s) of the highest-energy photon reported by LAT 77 s after the trigger time. In all the time windows including around the highest-energy photon were consistent with background only.\\

At the trigger time reported by Fermi-LAT,  GRB 160509A was at an elevation of  $\theta\approx 27^\circ.98$  and culminated at $33^\circ$ inside the HAWC's field of view \citep{2016GCN.19423....1L}.  Taking into account the sensitivity of HAWC to GRBs described in \cite{2012APh....35..641A}  for  a range of the zenith angle between $0.7\geq \cos\theta > 0.6$, an upper limit at  $100\,{\rm GeV}$ and a spectral index of $2$ could be derived as $\approx 1.14\times 10^{-4}\,{\rm mJy}$. In this case, the theoretical flux predicted  through eq. (\ref{ssc_ism}) is $\simeq 10^{-11}\,{\rm mJy}$,  which agrees with  the upper limit derived.\\

\section{Predicted Light Curve of GRB 180418A:  Thin-shell case} 

At 06:44:06.012 UTC on 18 April 2018, the Swift BAT instrument triggered and located GRB 180418A \citep{2018GCN.22646....1D}. The BAT light curve in the energy range of 15 -150 keV exhibited a single FRED-like pulse with a duration of $T_{90}=1.504\pm 0.380\,{\rm s}$.   At 06:44:06.28 UTC on 18 April 2018,  Fermi GBM triggered on GRB 180418A \citep{2018GCN.22656....1B}.   The light curve consisted of a single FRED-like peak similar to the BAT light curve, with a duration of  $T_{90}=2.56\pm0.20\,{\rm s}$, and a fluence of $(5.9\pm0.1) \times 10^{-7}\,{\rm erg\,cm^{-2}}$ measured in the energy range of 10-1000 keV. The Fermi-LAT instrument did not detect GRB 180418A.  Because of an observing constraint,  the Swift XRT and UVOT instruments could not begin observing this burst until 3081.4 s and 3086 s after the trigger time, respectively.  The TAROT and RATIR optical telescopes started observing GRB 180418A in several filters 28.0 and 120.6 s after the trigger time, respectively \citep{2019ApJ...881...12B}.\\

\cite{2019ApJ...881...12B} presented observations of GRB 180418A in $\gamma$-ray, X-ray, and optical bands suggesting that this burst may have been a sGRB. This burst exhibited a bright optical flare \cite[$\approx 14.2$ AB mag in the r band;][]{2019ApJ...881...12B}  peaking  between 28 and 90 s after the trigger time.  The early optical observations were interpreted as synchrotron RS model  in the thin-shell regime and in a uniform-density medium for $p=2.35\pm0.03$.  Taking into account the isotropic-equivalent kinetic energy of $E_{\rm K}=7.7\times 10^{50}\,{\rm erg}$ and a redshift of $z=0.5$, the values of the parameters found by \cite{2019ApJ...881...12B} are: the circumburst density  ($n=0.15\,{\rm cm^{-3}}$), the bulk Lorentz factor ($\Gamma=160$) and the microphysical parameters ($\epsilon_{\rm e_r}=0.1$ and $\epsilon_{\rm B_r}=0.2$).\\

\begin{figure}[h!]
{ \centering
\resizebox*{0.7\textwidth}{0.35\textheight}
{\includegraphics{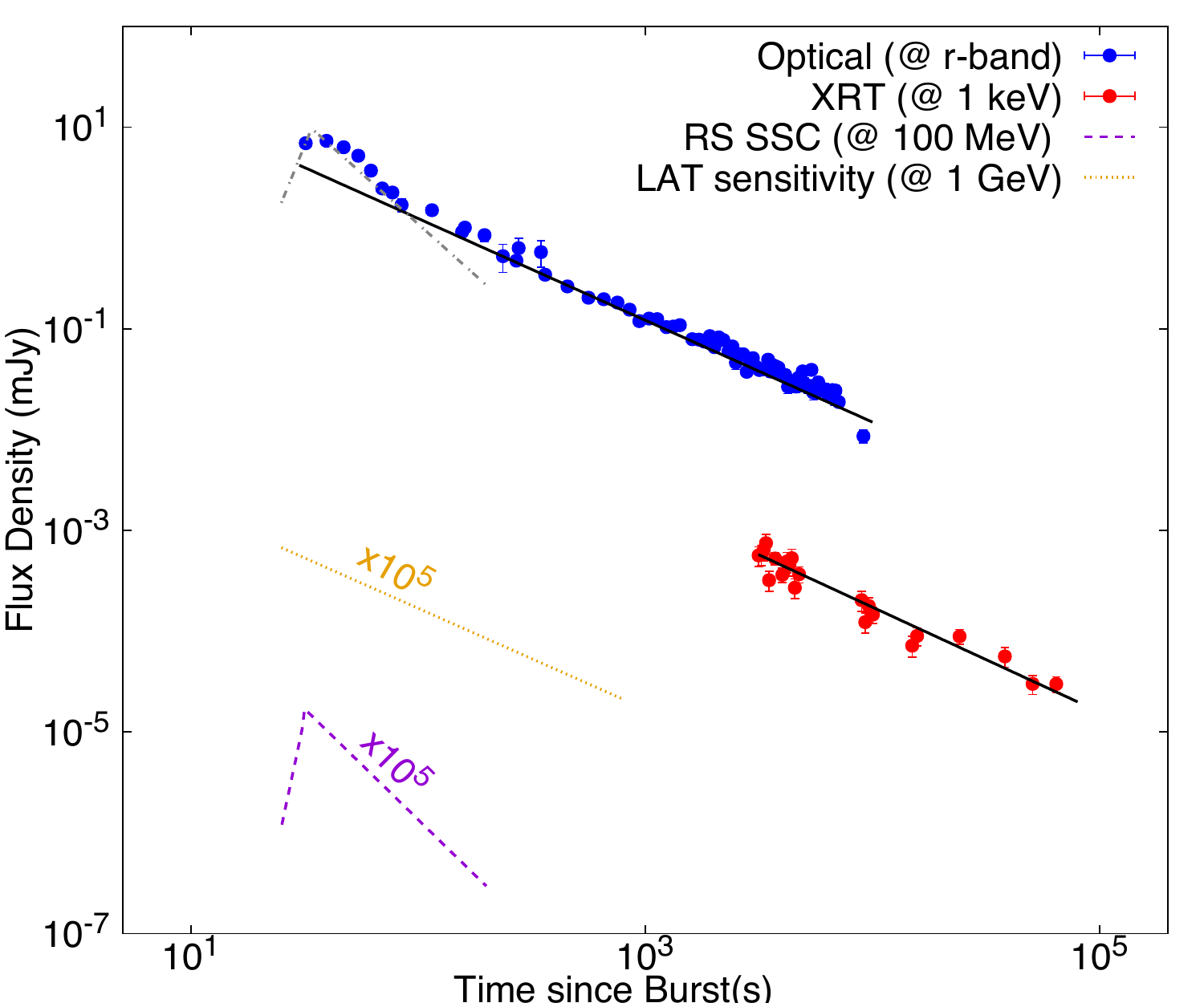}}
}
\caption{Multiwavelength light curves of GRB 180418A, with the best-fit curves.  The synchrotron FS model was used  to describe the X-ray and optical observations,  and the synchrotron  RS model  was used to describe the bright optical peak.  The theoretically predicted SSC emission in the thin-shell regime and the Fermi-LAT sensitivity at 1 GeV is  displayed in magenta and yellow, respectively.  Data points with the fit are taken from \cite{2019ApJ...881...12B}. }
 \label{fig4:multiwavelength GRB180418A}
\end{figure}

Using eqs. (\ref{slow_before_thin}) and (\ref{slow_after_thin}) and the parameters reported by  \cite{2019ApJ...881...12B}, we plot the theoretically predicted SSC emission from the RS evolving in the thin-shell regime and the Fermi-LAT sensitivity extrapolated at 1 GeV,  as shown in Figure \ref{fig4:multiwavelength GRB180418A}.  This figure shows the X-ray and optical observations with the best-fit curve for synchrotron RS emission\footnote{The derived quantities as the synchrotron break frequencies, fluxes and the evolution of the Lorentz factor are reported in \cite{2019ApJ...881...12B}.} and the predicted SSC emission.\footnote{The optical peak is well separated from the prompt emission, which is a hallmark of the thin-shell model.}  It shows that the SSC emission is around two orders of magnitude below the Fermi-LAT sensitivity, which agrees with the non-detections reported by this instrument. 

\section{Summary}

We have derived the SSC light curves from RS  in thick- and thin-shell regimes for a uniform-density medium and shown that this emission in the thick-shell case could describe the GeV flares exhibited in some interesting LAT-detected bursts \citep{2010MNRAS.403..926G,2017ApJ...843..114L}.   Since the shock crossing time is less than the prompt emission in the thick-shell regime, a bright RS SSC peak is expected at the beginning of the FS emission. By contrast, as  the shock crossing time is longer than the duration of the prompt emission in the thin-shell regime, a bright RS SSC peak, in this case, appears distinct from the prompt emission and can be expected during the FS emission.   The rise and decay indices of the RS SSC light curves are expected to be  $-\frac{p+3}{4} \lesssim \alpha_{\rm ris} \lesssim -\frac14$  and $\frac{29}{96} \lesssim \alpha_{\rm dec} \lesssim \frac{p+4}{2}$ for a thick-shell regime, and $-\frac{12p-7}{2} \lesssim \alpha_{\rm ris}\lesssim -\frac12$  and  $\frac{6}{35} \lesssim \alpha_{\rm dec} \lesssim \frac{p+4}{2}$ for a thin-shell regime, respectively.  We have shown that a bright RS SSC peak is expected when the microphysical parameter $\epsilon_{\rm e, r}$ is above 0.1 and suppressed when $\mathcal{R}_{\rm e} \ll 1$.\\

We have also investigated the nature of late-time breaks in GeV light curves, and interpret these as the passage of the synchrotron cooling frequency of the FS through the GeV band. This naturally occurs in a uniform, low-density environment. For 
more energetic bursts, a lower density is needed for this effect to be observed.   We have shown that the simultaneous presence of GeV flare and a break in the LAT light curve requires a low density and $\mathcal{R}_{\rm B}  \gg 1$,  suggesting that the outflow could be endowed with primordial magnetic fields in such cases.\\  

We emphasize that the FS closure relations are not expected to be satisfied when the LAT light curves are a superposition of SSC and synchrotron from RS and FS, respectively. Only when the RS SSC emission is suppressed or has decreased sufficiently so that it is negligible regarding the synchrotron FS emission, the closure relation can be satisfied in the LAT energy band. It is worth highlighting that, depending on the parameter values, GeV flare  RS  SSC emission could be hidden by longer-lasting FS synchrotron emission.\\

As a particular case for the thick-shell regime, we have studied the LAT observations of GRB 160509A, which exhibited a clear, bright peak at 20 s with a break at 316 s in the light curve.   With the values of the best-fit parameters, we inferred that the first photon with an energy of 51.9 GeV arriving at 76.5 s and the second one of 41.5 GeV arriving at 242 s after the trigger time is produced in the deceleration phase of the outflow and a different mechanism to the standard synchrotron model has to be invoked to interpret these VHE photons. We explicitly showed that the SSC FS emission generates these VHE photons.  The best-fit values of the microphysical parameters indicate that a magnetized outflow could explain the features exhibited in the light curves of GRB 160509A.\\
As an example of the thin-shell regime, we have predicted the light curves at 100 MeV for GRB 180418A with the parameters used to describe the X-ray and optical observations. As expected, the light curves at 100 MeV are below the Fermi-LAT sensitivity.

\acknowledgements
We thank  Xiang-Yu Wang, Alan Watson  and B.B Zhang for useful discussions and comments. NF  acknowledges  the support  from UNAM-DGAPA-PAPIIT  through  grants  IA102019 and IN107518.  RBD  acknowledges support  from  National Science Foundation (NSF) under grant 1816694.  M.G.D. acknowledges the support from the American Astronomical Society Chretienne Fellowship and from MINIATURA2 grant, 2018/02/X/ST9/03673.
%
%\bibliography{Bib_GRB160509A}
%\bibliographystyle{apa}

\addcontentsline{toc}{chapter}{Bibliography}

\newpage
\begin{table}
\centering \renewcommand{\arraystretch}{2}\addtolength{\tabcolsep}{3pt}
\caption{The best-fit values found from the LAT light curve of GRB 160509A}
\label{table1}
\begin{tabular}{l  c  c  c }
 \hline \hline
\scriptsize{LAT} &  \scriptsize{Parameter}  &\hspace{0.5cm}   \scriptsize{Best-fit value}    & \hspace{0.5cm} \scriptsize{ $\chi^2$/ndf} \\ 
\hline \hline

\scriptsize{GeV flare} \\
  	        &  \scriptsize{$\alpha_{\rm L, ris}$}  &\hspace{0.5cm} \scriptsize{$1.25\pm0.21$}		&\hspace{0.5cm}  \scriptsize{$1.22$}\\ %\cdashline{1-4}
                  	        &  \scriptsize{$\alpha_{\rm L, dec}$}  &\hspace{0.5cm} \scriptsize{$3.14\pm0.65$}		&\hspace{0.5cm}  \scriptsize{$$}\\ %\cdashline{1-4}
                  	        &  \scriptsize{$t_{\rm 0}$ (s)}  &\hspace{0.5cm} \scriptsize{$8.4\pm0.5$}		&\hspace{0.5cm}  \scriptsize{$$}\\ \cdashline{1-4}
\scriptsize{Extended emission}   	\\
\scriptsize{}        &  \scriptsize{$\alpha_{L,1}$}  &\hspace{0.5cm} \scriptsize{$0.87\pm0.18$}		&\hspace{0.5cm}  \scriptsize{$1.27$}\\	
   		 	        &  \scriptsize{$\alpha_{L,2}$}  &\hspace{0.5cm} \scriptsize{$1.26\pm0.21$}		&\hspace{0.5cm}  \scriptsize{$$}\\	%\cdashline{1-4}   
   		 	        & \scriptsize{$t_{\rm L, br}$ (s)}  &\hspace{0.5cm} \scriptsize{$316.2\pm61.2$}		&\hspace{0.5cm}  \scriptsize{$$}\\	%\cdashline{1-4}  
			        
\hline \hline
\end{tabular}
\end{table}
\begin{table}[h!]
\centering \renewcommand{\arraystretch}{2}\addtolength{\tabcolsep}{3pt}
\caption{The best-fit values found from the XRT light curve of GRB 160509A}
\label{table2}
\begin{tabular}{c  c  c  c }
 \hline \hline
\scriptsize{X-rays} &\hspace{0.5cm}   \scriptsize{Time interval}  &\hspace{0.5cm}   \scriptsize{Index}    & \hspace{0.5cm} \scriptsize{ $\chi^2$/ndf} \\ 
\scriptsize{} & \hspace{0.5cm} \scriptsize{(s)} & \hspace{0.5cm}  \scriptsize{($\alpha_{\rm X}$)}   &   \\ 
\hline \hline
\scriptsize{II}   	        & \hspace{0.5cm} \scriptsize{$(0.35 - 5.1)\times 10^4$}  &\hspace{0.5cm} \scriptsize{$0.50\pm0.12$}		&\hspace{0.5cm}  \scriptsize{$1.22$}\\\cdashline{1-4}
\scriptsize{III}   	        & \hspace{0.5cm} \scriptsize{$(0.51 - 4.28) \times10^5$}  &\hspace{0.5cm} \scriptsize{$1.23\pm0.13$}		&\hspace{0.5cm}  \scriptsize{$1.11$}\\\cdashline{1-4}
\scriptsize{IV}   	        & \hspace{0.5cm} \scriptsize{$\geq 4.28\times10^5$ }  &\hspace{0.5cm} \scriptsize{$2.06\pm0.3$}		&\hspace{0.5cm}  \scriptsize{$1.51$}\\
\\
\hline \hline
\end{tabular}
\end{table}
%
%
%\newpage
\begin{table}[h!]
\centering \renewcommand{\arraystretch}{2}\addtolength{\tabcolsep}{3pt}
\caption{Median values of parameters of GRB 160509A found with symmetrical quantiles.  The external-shock model is used to constrain the values of the parameters.}
\label{table3}
\begin{tabular}{ l  c c}
\hline
\hline
{\large   Parameters}	& {\large  Median}   	& {\large $\hat{R}$}	 \\ 
\hline \hline
\\
\small{$E_{\rm K}\,(10^{53}\,{\rm erg})$}	\hspace{1.5cm}&     \small{$6.977^{+0.801}_{-0.802}$} & 1.000 \\
\small{$n\,(10^{-4}\, {\rm cm^{-3}})$}	\hspace{1.5cm}&     \small{$4.554^{+1.128}_{-1.121}$} & 0.999	\\
\small{$p$}	\hspace{1.5cm}&     \small{$2.400^{+0.079}_{-0.081}$} &	1.000 \\
\small{$\epsilon_{\rm B_f}\,(10^{-5})$}	\hspace{1.5cm}&     \small{$4.105^{+0.938}_{-0.918}$} &	0.999  \\
\small{$\epsilon_{\rm B_r}\,(10^{-1})$}	\hspace{1.5cm}&     \small{$1.100^{+0.099}_{-0.100}$} &	0.999  \\
\small{$\epsilon_{\rm e_f}\,(10^{-2})$}	\hspace{1.5cm}&     \small{$3.101^{+0.305}_{-0.305}$} &	0.999	 \\
\small{$\epsilon_{\rm e_r}\,(10^{-1})$}	\hspace{1.5cm}&     \small{$8.000^{+0.080}_{-0.079}$} &	0.999	 \\
\small{$\alpha_e\,(10^{-1})$}	\hspace{1.5cm}&     \small{$-3.199^{+0.366}_{-0.375}$} &	1.000 \\
\small{$\alpha_B\,$}	\hspace{1.5cm}&     \small{$1.401^{+0.136}_{-0.142}$} &	0.999	 \\
\hline
\end{tabular}
\end{table}
\end{document}